\documentclass[%
aps,
prd,
amssymb,
amsmath,
amsfonts,
eqsecnum,
showpacs,
nofootinbib,
floatfix,
twocolumn,
twoside,
a4paper,
superscriptaddress,
longbibliography
]{revtex4-1}

\usepackage{amsmath, latexsym}
\usepackage{amssymb}

\usepackage[T3,T1]{fontenc}
\usepackage{mathrsfs} 
\usepackage{bm} 
\makeatletter
\@ifclassloaded{beamer}
  { 
    \typeout{UsePackages: Detected beamer}
    \usepackage{tgheros}
    
  }
  { 
    \typeout{UsePackages: Did not detect beamer}
    \ifx\asybeamer\undefined 
    \typeout{UsePackages: Detected article}
    \usepackage{tgtermes} 
    \else 
    \typeout{Fonts: Detected asy for beamer}
    \usepackage{tgheros}
    
    \fi
  }
\usepackage{microtype} 
\def\MT@register@subst@font{
  \MT@exp@one@n\MT@in@clist\font@name\MT@font@list
  \ifMT@inlist@\else\xdef\MT@font@list{\MT@font@list\font@name,}\fi}
\makeatother

\usepackage{graphicx} %
\usepackage[x11names,svgnames,rgb]{xcolor} %
\usepackage{xspace}
\usepackage{braket}
\usepackage{accents}
\usepackage{siunitx}
\usepackage{mathtools}
\DeclareSymbolFontAlphabet{\mathrm}{operators}

\definecolor{CiteColor}{rgb}{0.18039, 0.18824, 0.57255}
\definecolor{UrlColor} {rgb}{0.741, 0.173, 0.000}
\definecolor{DarkUrlColor} {rgb}{0.500, 0.110, 0.000}
\definecolor{LinkColor}{rgb}{0.25098, 0.47843, 0.04706}

\makeatletter %
\newcommand{\ShowFont}{%
  \typeout{The main font is \f@encoding \space \f@family \space %
    \f@series \space \f@shape \space at \f@size pt.}%
  \typeout{The math font sizes are \tf@size pt (main), \sf@size pt %
    (script), and \ssf@size pt (scriptscript).}%
  \typeout{The linewidth is \the\linewidth}} %
\makeatother %

\usepackage{xargs}

\usepackage{graphicx}
\usepackage{dcolumn}
\usepackage{bm}
\usepackage{tabularx}
\usepackage{multirow}
\usepackage{capt-of}
\usepackage{color}
\usepackage{url}
\usepackage[utf8]{inputenc}
\usepackage{placeins}
\usepackage{soul}

\usepackage[nolist,nohyperlinks]{acronym}
\usepackage{xspace}
\usepackage[colorlinks]{hyperref}

\usepackage{subfigure}
\usepackage{booktabs}
\usepackage{orcidlink}

\DeclareMathAlphabet{\mathbfsf}{\encodingdefault}{\sfdefault}{bx}{sl}

\newcommand{\be}{\begin{equation}}
\newcommand{\ee}{\end{equation}}
\newcommand{\bea}{\begin{eqnarray}}
\newcommand{\eea}{\end{eqnarray}}

\definecolor{light-gray}{gray}{0.95}

\definecolor{dodgerblue}{HTML}{1E90FF}
\definecolor{viennared}{HTML}{DA0A14}
\definecolor{ctorange}{HTML}{FF6C0C}
\definecolor{granadagreen}{HTML}{078931}
\definecolor{wales}{HTML}{ff0038}
\definecolor{valenciacfred}{HTML}{ee3524}
\definecolor{barcelonafcgold}{HTML}{edbb00}
\definecolor{jam}{HTML}{A50B5E}
\definecolor{austriawien}{HTML}{441678}

\AtBeginDocument{%
  \hypersetup{
    citecolor=dodgerblue,
    linkcolor=dodgerblue,   
    urlcolor=dodgerblue}}

\newcommand{\UIB}{Departament de F\'isica, Universitat de les Illes Balears, IAC3 -- IEEC, Crta. Valldemossa km 7.5, E-07122 Palma, Spain}

\newcommand{\ICE}
{Institut de Ci\`encies de l'Espai (ICE, CSIC), Campus UAB, Carrer de Can Magrans s/n, 08193 Cerdanyola del Vall\`es, Spain}

\newcommand{\Marseille}
{Aix-Marseille Universit\'e, Universit\'e de Toulon, CNRS, CPT, Marseille, France}

\allowdisplaybreaks[1]

\begin{document}

\title[ML]
{Impact of the $(\ell=2,m=0)$ spherical harmonic mode with memory on parameter estimation for ground-based detectors}


\author{Maria Rossell\'o-Sastre\,\orcidlink{0000-0002-3341-3480}}
\affiliation{\UIB}

\author{Sascha Husa\,\orcidlink{0000-0002-0445-1971}}
\affiliation{\ICE}
\affiliation{\UIB}

\author{Sayantani Bera\,\orcidlink{0000-0003-0907-6098}}
\affiliation{\Marseille}
\affiliation{\UIB}

\author{Yumeng Xu\,\orcidlink{0000-0001-8697-3505}}
\affiliation{\UIB}

\date{\today}

\begin{abstract}
We recently presented an efficient and accurate waveform model for the $(2,0)$ spherical harmonic mode including both the displacement memory contribution and the ringdown oscillations for aligned-spin binary black holes in quasi-circular orbits. The model we developed is constructed in time domain and implemented within the computationally efficient {\tt IMRPhenomTHM} waveform model. In this article, we employ it to perform in-depth parameter estimation studies for future ground-based detectors, specifically considering LIGO A$^{\#}$, Cosmic Explorer, and the Einstein Telescope, combining them in different detector networks. While previous studies have reviewed the impact of the memory contribution in parameter estimation, we assess the effect of incorporating the complete mode in the analysis on the posterior estimation of source parameters, performing zero-noise injections of high signal-to-noise ratio signals. We investigate the impact of this mode on the distance-inclination degeneracy and compare its impact in edge-on and face-on configurations. We find that including this mode helps mitigate biases in the estimation of individual spin components, which may otherwise arise when the mode is neglected. 
\end{abstract}


\maketitle

\section{Introduction}
\label{sec:Introduction}
With the ongoing enhancement of gravitational-wave detector sensitivities and the development of more advanced observatories, the demand for more accurate waveform models is increasing. Accurate models are essential for the reliable inference of source properties and the mitigation of biases in parameter estimation. Accordingly, the inclusion of subdominant physical effects, which capture previously neglected aspects of the waveform, is crucial for improving model accuracy.

In this context, we recently developed a model for the $(\ell=2,m=0)$ spherical harmonic mode of spin-aligned binary black hole (BBH) systems in quasicircular orbits \cite{PhysRevD.110.084074}, which was the last unmodeled component of the quadrupole $(\ell=2)$ modes in the IMRPhenom family of waveform models. This mode includes, apart from the quasinormal mode oscillations in the ringdown, the main contribution of the displacement memory.  Constructed following the strategies employed by phenomenological waveform models \cite{PhysRevLett.113.151101, PhysRevD.100.024059, PhysRevD.101.024056, PhysRevD.103.104056, PhysRevD.93.044006, PhysRevD.93.044007, PhysRevLett.120.161102, PhysRevD.102.064001, PhysRevD.102.064002, estelles2020time, PhysRevD.103.124060, Estell_s_2022}, it has been implemented as an extension of {\tt IMRPhenomTHM} within {\tt LALsuite} \cite{lalsuite} and is also available through the {\tt phenomxpy} Python package \cite{phenomxpy}, enabling efficient application in parameter estimation studies.

Recent developments have addressed the modeling of this spherical harmonic mode. In addition to the NRSurrogate model {\tt NRHybSur3dq8\_CCE} \cite{PhysRevD.108.064027}, which incorporates memory effects, the $m=0$ modes have recently been included in {\tt TEOBResumS-GIOTTO} \cite{Albanesi:2024fts}; and a model both in time-domain and frequency-domain for the $(2,0)$ mode from nonspinning binaries has been presented in \cite{elhashash2025waveformmodelsgravitationalwavememory}. Furthermore, the memory contribution can be computed and added to different waveform models employing the {\tt GWMemory} package \cite{PhysRevD.98.064031}.

In our previous analysis \cite{PhysRevD.110.084074}, we found that our model is less accurate than {\tt NRHybSur3dq8\_CCE}, but it spans a broader region of the parameter space and delivers significantly faster evaluation times, making it particularly well suited for parameter estimation applications and for the generation of long-duration waveforms required by third-generation (3G) detectors.

In Ref.~\cite{PhysRevD.110.084074}, preliminary results concerning the detectability of the $(2,0)$ mode were presented by computing the signal-to-noise ratio (SNR) under various Advanced LIGO sensitivity scenarios. In the present work, we extend this analysis by performing parameter estimation studies that account for the planned upgrades of the Advanced LIGO detectors \cite{LIGOScientific:2014pky} and the sensitivities of future observatories, including the Einstein Telescope (ET) \cite{Abac:2025saz} and Cosmic Explorer (CE) \cite{CE}. We focus on future ground-based detectors because of their increased sensitivity, especially at low frequencies, which makes the subtle effects of the mode measurable and potentially bias-inducing, whereas current LIGO-Virgo-KAGRA (LVK) detectors lack the sensitivity to be significantly affected by this mode, at least for the analysis of single events.

The impact of the gravitational-wave memory in parameter estimation has already been studied in previous works. For instance, in Ref.~\cite{Xu:2024ybt}, the potential of breaking the distance-inclination degeneracy is explored using the displacement memory for the A$^{\#}$ sensitivity of the Advanced LIGO detector. In Ref.~\cite{Sun:2024nut}, the authors discuss the effect of the memory in parameter estimation for the future detector TianQin. However, these studies focus solely on the memory contribution and do not account for the full $(2,0)$ spherical harmonic mode, which also includes the oscillatory component. To date, parameter estimation analyses incorporating the full $(2,0)$ mode have not been conducted due to the absence of an efficient waveform model capable of describing it. Moreover, all the subdominant harmonics present in {\tt IMRPhenomTHM} are included in this work for the first time in an analysis of this type.

The objective of this work is to utilize the recently developed model for the full $(2,0)$ spherical harmonic mode to assess how this mode can enhance parameter estimation results. In Ref.~\cite{PhysRevD.110.084074}, it was demonstrated that the primary contribution to the SNR arises from the oscillatory component of the mode, although the memory component can provide a significant enhancement for specific regions of parameter space.

This article is organized as follows. In Sec.~\ref{sec:memory} we outline the properties of the displacement memory as well as a method to compute its contribution to the $(2,0)$ spherical harmonic. In Sec.~\ref{sec:tech} we give some details about the model used in this work and summarize the techniques employed in the parameter estimation studies we conduct. In Sec.~\ref{sec:LIGO} we present the results for the LIGO detector network and in Sec.~\ref{sec:3G}, for the networks of detectors combining LIGO with ET and CE. In Sec.~\ref{sec:conclusions} we summarize the main conclusions of our work. Finally, in Appendix \ref{app:cornerplots} and Appendix \ref{app:cornerplots_ET_CE} we present the full corner plots for all the parameter estimation injections performed in this work.

\section{Displacement memory}
\label{sec:memory}
Gravitational-wave memory is a nonlinear, nonoscillatory phenomenon sourced by the energy flux of the gravitational waves themselves, resulting in a permanent displacement in the spacetime that persists once the waves have passed. Although predicted, it has not yet been directly observed; its detection would constitute a fundamental test of general relativity. Based on the Bondi–Metzner–Sachs (BMS) balance laws, three distinct types of memory have been identified: displacement (or ordinary) memory, spin memory, and center-of-mass memory. In this work, we focus exclusively on displacement memory, as it directly appears in the gravitational-wave strain and represents the dominant contribution for binary black hole (BBH) mergers, particularly in the $(\ell=2,m=0)$ spherical harmonic mode.

The dependence of displacement memory on intrinsic binary parameters is well understood: it is maximized for equal-mass, spin-aligned systems, which produce the largest gravitational-wave energy flux (see Fig.~3 in Ref~\cite{Xu:2024ybt}). Furthermore, the $(2,0)$ spherical harmonic exhibits a distinct inclination dependence, being maximal for edge-on systems and vanishing for face-on systems. This behavior contrasts with the dominant $(2,\pm2)$ modes, which are strongest for face-on orientations.

The memory contribution can be computed from the energy flux term that appears in the supermomentum balance law \cite{Mitman:2020bjf},
\begin{equation}
\label{je}
    J_{\varepsilon}=\frac{1}{2}\bar{\eth}^2\mathfrak{D}^{-1}\left[\frac{1}{4}\int_{u_1}^u\left|\dot{h}\right|^2 du\right]+\alpha(\theta,\phi).
\end{equation}
After decomposing the strain into the basis of spin-weighted spherical harmonics,
\begin{equation}
    h(t,r,\theta,\phi)=h_+-ih_{\times}=\frac{1}{r}\sum_{\ell=2}^{\infty}\sum_{m=-\ell}^{m=\ell}h_{\ell,m}(t)\;^{-2}Y_{\ell,m}(\theta,\phi)
\end{equation}
and applying the corresponding operators, one can get to a simple analytic expression for the memory in the $(2,0)$ mode:
\begin{equation}
\begin{split}
\label{h20memhm}
h_{2,0}^{\text{(mem)}}&=\frac{1}{7}\sqrt{\frac{5}{6\pi}}\int_{u_1}^u\left|\dot{h}_{2,2}\right|^2\;du-\frac{1}{14}\sqrt{\frac{5}{6\pi}}\int_{u_1}^u\left|\dot{h}_{2,1}\right|^2\;du\\
&+\frac{5}{2\sqrt{42\pi}}\int_{u_1}^u\left(\dot{h}_{2,2}^{\text{Re}}\dot{h}_{3,2}^{\text{Re}}+\dot{h}_{2,2}^{\text{Im}}\dot{h}_{3,2}^{\text{Im}}\right)du\\
&-\frac{2}{11}\sqrt{\frac{2}{15\pi}}\int_{u_1}^u\left|\dot{h}_{4,4}\right|^2\;du.
\end{split}
\end{equation}
In this case, we decompose the strain considering only the modes $(\ell,m)=(2,\pm2),(2,\pm1),(3,\pm2),(3,\pm3),(4,\pm4)$ as they provide the most significant contributions. The details of this calculation can be found in \cite{PhysRevD.110.084074}.

\section{Technical preliminaries}
\label{sec:tech}
In this section, we outline the technical aspects of the waveform model employed and the setup used for parameter estimation studies.

\subsection{The $\mathbf{(\ell=2,m=0)}$ spherical harmonic mode within the IMRPhenomTHM waveform model}
\label{sec:THM}
The {\tt IMRPhenomTHM} model \cite{estelles2020time} is a time-domain inspiral-merger-ringdown (IMR) phenomenological model from the Phenom family of waveform models that describes spin-aligned BBH mergers in quasicircular orbits. In addition to the dominant quadrupolar modes $(2,\pm2)$, it also incorporates the $(2,\pm1), (3,\pm3), (4,\pm4)$ and $(5,\pm5)$ subdominant spherical harmonics. We have extended this model by incorporating our recently developed description of the $(2,0)$ mode. We refer to this extended version of the model as {\tt IMRPhenomTHM+(2,0)} \cite{PhysRevD.110.084074}.

The $(2,0)$ mode in this extended model includes both the nonoscillatory memory contribution and the oscillatory component associated with the quasinormal ringdown. Each contribution is modeled independently by fitting analytical ansätze. This construction yields computationally efficient, closed-form expressions suitable for rapid evaluation in parameter estimation studies.

The relative contributions of the oscillatory and memory components depend on the mass ratio $1/q=m_1/m_2\geq1$ and the dimensionless spin components $\chi_{1z}$ and $\chi_{2z}$ of the individual black holes. To illustrate this behavior, in Fig.~\ref{fig:amplitudes}, we present an example of equal-spin configurations in each panel: high negative spins, nonspinning, and high positive spins. The plots illustrate how the maximum amplitude of the different components of the $(2,0)$ mode varies as a function of the mass ratio. We select spin magnitudes $|\chi_{1z}|=|\chi_{2z}|=0.7$ since, although the model can be used for higher values of the spin amplitude, we know that it loses accuracy in this region of the parameter space. Unequal spin cases are excluded from this plot, as the selected configurations already capture the representative waveform features. As it is an $m=0$ mode, it has only one polarization; therefore, the real part of the mode corresponds to the full amplitude. Across the three panels, we observe that the memory amplitude is maximum for equal-mass systems and increases with higher positive spin magnitudes. In contrast, the amplitude of the oscillatory component only approaches that of the memory for systems with low and high negative spins with $1/q$ of approximately $10$ and $3$, respectively. For the high negative spin case, there is a crossover at around $1/q=3.5$, which indicates that beyond this, the oscillatory part becomes more dominant compared to the memory contribution. The exact crossover point depends on the spin components of the black holes. The oscillatory amplitude remains relatively constant as the mass ratio changes for nonspinning and high positive spin systems. As a result, the overall amplitude of the $(2,0)$ mode is dominated by the memory contribution, particularly for systems with high positive spins, where the oscillatory component becomes nearly negligible.

The detectability of the oscillatory and memory components depends critically on the frequency sensitivity of gravitational-wave detectors. Current LVK detectors lack sufficient low-frequency sensitivity to directly observe the zero-frequency memory contribution. In addition, noise-weighting the signal by the detectors' power spectral density (PSD) further suppresses the low-frequency components, where the memory effect is strongest. Consequently, the whitened memory signal, originally a step-like feature in the time domain, is transformed into a decaying oscillatory pattern. As a result, the observed $(2,0)$ mode in current detectors will predominantly exhibit the ringdown oscillations, supplemented by oscillatory features associated with the whitened memory. In contrast, future detectors such as ET, with improved low-frequency sensitivity (see Fig.~\ref{fig:PSDs}), will enable the direct detection of a larger fraction of the memory contribution.

\begin{widetext}
\begin{center}
\begin{figure}[htp]
\includegraphics[width=1\textwidth]{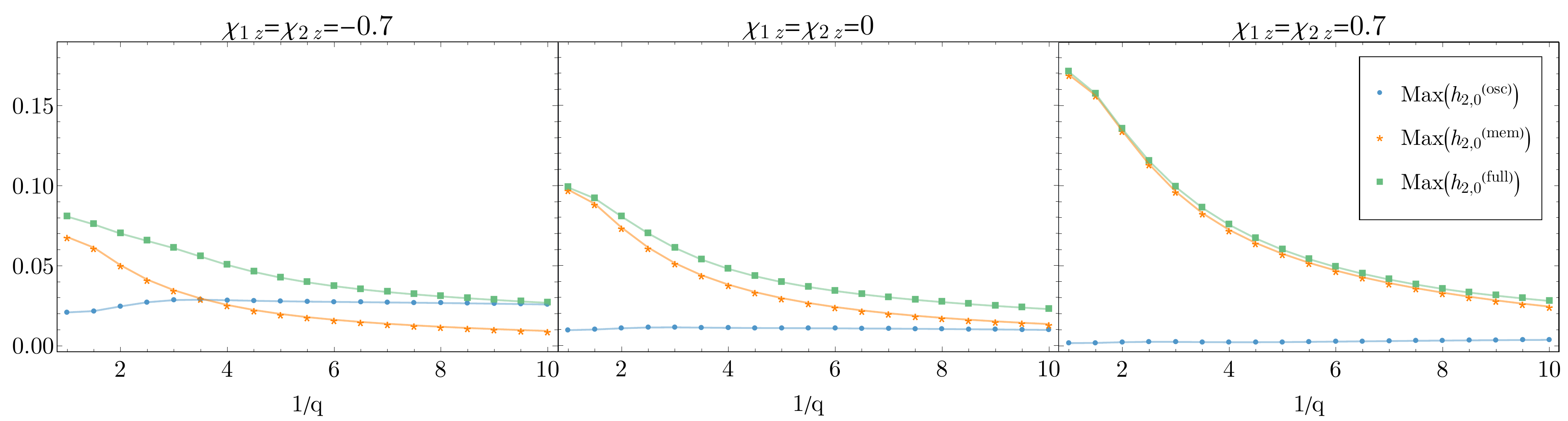}
    \caption{Dependence on the mass ratio of the maximum of the strain amplitude (in geometric units, $G=c=1$) for equal-spin systems of the full $(2,0)$ spherical harmonic (green squares), the oscillatory component (blue circles) and the memory component (orange stars) of this mode. The left panel corresponds to a configuration with high negative spins, the middle panel to a nonspinning system, and the right panel to a configuration with high positive spins.}
    \label{fig:amplitudes}
\end{figure}
\end{center}
\end{widetext}

\subsection{Notation and conventions}
In the following sections, we refer to various source parameters, which we first introduce here. In the case of the mass parameters, all quantities are given in the detector frame. We denote the masses of the individual components by $m_1$ and $m_2$, then the chirp mass is defined as $\mathcal{M}=\frac{(m_1m_2)^{3/5}}{(m_1+m_2)^{1/5}}$ and the mass ratio as $q=m_2/m_1\leq1$. Since we restrict our analysis to nonprecessing systems, we only consider the $z$ components of the dimensionless spin vectors, as they are the ones aligned with the orbital angular momentum, and we denote them as $\chi_{1z}$ and $\chi_{2z}$. The angle between the angular momentum vector and the line of sight between the binary and the detector is referred to by $\theta_{\text{JN}}$, and the luminosity distance is represented as $d_L$. The sky location of the binary is specified using equatorial coordinates, namely right ascension denoted by $\alpha$, and declination denoted by $\delta$. The reference phase and polarization angle are denoted by $\phi$ and $\psi$, respectively. Finally, the coalescence time is indicated by $t_c$.

\subsection{Bayesian Parameter Estimation}
\label{sec:pe}
We briefly outline the basics of parameter estimation. According to Bayes' theorem, the posterior distribution of parameters given some observed data is
\begin{equation}
    \label{bayes}
    p(\Theta|d)=\frac{\mathcal{L}(d|\Theta)\pi(\Theta)}{\mathcal{Z}},
\end{equation}
where $\mathcal{L}(d|\Theta)$ is the likelihood of the data $d$ given the parameters $\Theta$, $\pi(\Theta)$ is the prior distribution of the parameters, and $\mathcal{Z}$ is a normalization factor given by
\begin{equation}
    \label{normfactor}
    \mathcal{Z}=\int d\Theta \mathcal{L}(d|\Theta)\pi(\Theta).
\end{equation}
The prior incorporates the previous knowledge about the parameters before carrying out any measurement, and the likelihood is evaluated using the data and the chosen waveform model. The expression for computing the likelihood in Eq.~(\ref{bayes}), assuming Gaussian noise, is given by
\begin{equation}
\label{likelihood}
    \mathcal{L}(d|\Theta)=\exp\left(-\frac{1}{2}\langle h_{\text{inj}}-h(\Theta)| h_{\text{inj}}-h(\Theta) \rangle\right),
\end{equation}
where $\langle \cdot | \cdot \rangle$ denotes the noise-weighted inner product and $h_{\text{inj}}$ is the injected signal. The inner product for two general waveforms $g(t)$ and $h(t)$, is defined in the frequency domain as
\begin{equation}
\label{innerprod}
    \langle g|h\rangle= 4 \text{Re}\left[\int_{f_{\text{min}}}^{f_{\text{max}}}\frac{\tilde{g}(f)\tilde{h}^*(f)}{S_n(f)}df\right],
\end{equation}
where $S_n(f)$ is the one-sided PSD of the detector noise.

Posterior sampling is performed using the Bilby software library \cite{bilby_paper} with {\tt bilby\_pipe} \cite{bilby_pipe_paper} and the nested sampling algorithm {\tt dynesty} \cite{dynesty}. In Tab.~\ref{tab:priors}, we summarize the priors used for the parameters we sample. Regarding the mass components, the individual component masses are uniformly distributed. Instead of directly applying a uniform prior to $q$ and $\mathcal{M}$, the samples for $m_{1,2}$ are drawn from a uniform distribution within the range stated in the constraint prior. For the luminosity distance, we use the Uniform Source Frame prior, which is uniform in comoving volume and source frame time. 

\begin{table}[h!]
\centering
\begin{tabular}{ccccc}
\toprule
\multicolumn{1}{l}{\textbf{Variable}} & \multicolumn{1}{l}{\textbf{Unit}} & \textbf{Prior}        & \textbf{Minimum} & \textbf{Maximum} \\ \hline
$m_{1,2}$                             & $M_\odot$                         & Constraint               & 10                & 100              \\
$q$                                   & -                                 & Uniform in Components & 0.125            & 1                \\
$\mathcal{M}$                         & $M_\odot$                         & Uniform in Components & 15               & 100              \\
$\chi_{1z,2z}$                        & -                                 & Aligned Spin               & -0.99                & 0.99             \\
$d_L$                                 & Mpc                               & Uniform Source Frame  & 20               & 1000             \\
$\theta_{\text{JN}}$                  & rad                               & Sin                   & 0                & $\pi$            \\
$\phi$                              & rad                               & Uniform               & 0                & $2\pi$           \\
$\delta$                              & rad                               & Cos                   & $-\pi/2$         & $\pi/2$          \\
$\alpha$                              & rad                               & Uniform               & 0                & $2\pi$           \\
$\psi$                                & rad                               & Uniform               & 0                & $\pi$            \\
\bottomrule
\end{tabular}
\caption{Prior distributions for the sampled parameters used in the injections.}
\label{tab:priors}
\end{table}

We generate injected signals using the {\tt IMRPhenomTHM+(2,0)} model as implemented in {\tt LALSuite}. We then recover parameters using both the same model (including the $(2,0)$ mode) and the {\tt IMRPhenomTHM} model (excluding the $(2,0)$ mode), enabling us to quantify the biases introduced when the mode is omitted. We inject and recover using the same waveform model in order to be sure that the biases that we find are due to the absence of the $(2,0)$ mode rather than to waveform systematics that could be found when using a different model.

\subsection{Fourier transform of the signals}
Since the $(2,0)$ mode includes a memory component, the waveform exhibits a nonzero offset after ringdown, breaking the periodicity assumption typically used in the discrete Fourier transform (DFT). Directly applying a DFT to such nonperiodic signals introduces artifacts in the frequency domain. To address this, several methods have been developed, see e.g. \cite{PhysRevD.110.064049,elhashash2025waveformmodelsgravitationalwavememory}. In this work, we employ the {\tt gw-foutstep} package \cite{PhysRevD.110.124026},  which produces an artifact-free Fourier transform suitable for signals with memory. This method is specified in Bilby through the {\tt frequency-domain-source-model} configuration option.

This must also be considered when calculating the SNR of the signals, as it is computed in the frequency domain. The optimal SNR for a signal $h(t)$ in a detector characterized by the one-sided PSD $S_n(f)$ is given by
\begin{equation}
    \rho=\sqrt{\langle h|h\rangle}=\sqrt{4\int_0^{\infty}\frac{|\tilde{h}(f)|^2}{S_n(f)}df}.
\end{equation}
When considering a network of multiple detectors (e.g. both LIGO Livingston and LIGO Hanford detectors) instead of a single detector, the joint SNR of the network is computed by summing the squared SNRs of the individual observatories,
\begin{equation}
    \rho_{\text{net}}=\sqrt{\sum_i \rho_i^2}.
\end{equation}

\begin{figure}[htp]
\includegraphics[width=0.85\columnwidth]{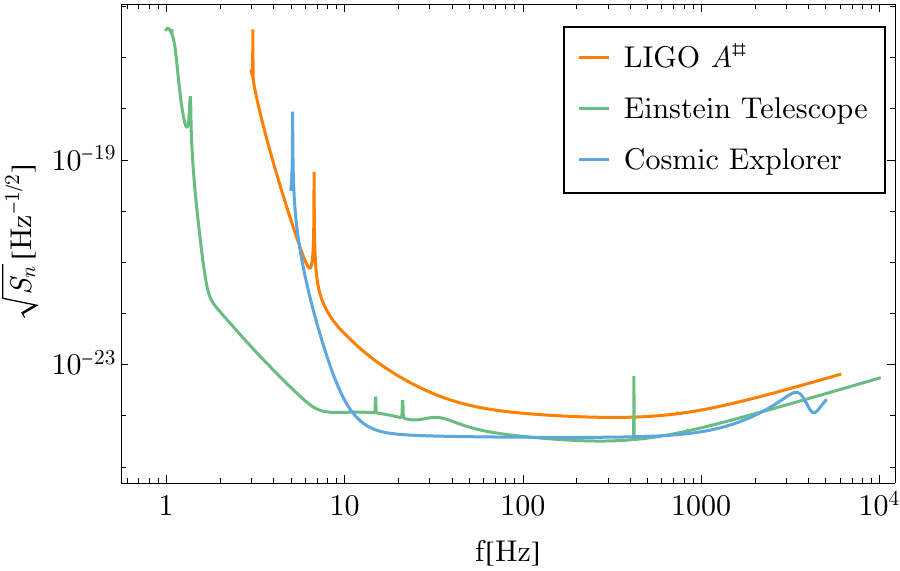}
    \caption{Sensitivity curves of LIGO A$^{\#}$, Einstein Telescope (ET-D) and Cosmic Explorer (CE1) detectors used in the injections, taken from \cite{Sensitivitycurves}.}
    \label{fig:PSDs}
\end{figure}

\section{Injections with LIGO A$^{\#}$ sensitivity}
\label{sec:LIGO}

We begin by performing a series of zero-noise injections into the LIGO Livingston (L1) and LIGO Hanford (H1) observatories \cite{LIGOScientific:2014pky} with the A$^{\#}$ design sensitivity, represented by the orange curve in Fig.~\ref{fig:PSDs}, taken from \cite{Sensitivitycurves}. In Tab.~\ref{tab:injections} we collect the parameters of the set of eight injections for which we present the results. Across different cases, we vary the mass ratio and chirp mass, leading to different values of the individual masses. We set $q=0.970$ rather than exactly $q=1$ in order to have a contribution from odd higher modes when we set equal-spin configurations. We select different combinations of the spin parameters, from negative to high positive values including a nonspinning case. We choose low values of the luminosity distance (which are still consistent with GW observations) in order to have considerably high values of the SNR, which are reduced when considering edge-on binaries with respect to their face-on equivalents. Since the $(2,0)$ is a subdominant mode - typically one or two orders of magnitude smaller than the dominant $(2,\pm2)$ mode, depending on the specific case - it is necessary to consider signals with high SNRs to ensure that the influence of the $(2,0)$ mode on the parameter estimation is sufficiently significant.

\begin{table}[h!]
\centering
\resizebox{\columnwidth}{!}{ 
\begin{tabular}{lcccccccc}
\toprule
                                         & $\bm{m_1[M_{\odot}]}$ & $\bm{m_2[M_{\odot}]}$ & $\bm{q}$ & $\bm{\mathcal{M}[M_{\odot}]}$ & $\bm{\chi_{1z}}$ & $\bm{\chi_{2z}}$ & $\bm{d_L}$\textbf{[Mpc]} & $\bm{\theta_{\text{\textbf{JN}}}}$\textbf{[rad]} \\ \hline
\textbf{Injection 1}                     & 50.8                  & 49.2                  & 0.970    & 43.5                          & 0.200            & 0.200            & 200                                           & $\pi/2$                                                               \\
\textbf{Injection 2}                     & 50.8                  & 49.2                  & 0.970    & 43.5                          & 0.200            & 0.200            & 200                                           & 0                                                                     \\
\textbf{Injection 3}                     & 35.5                  & 34.5                  & 0.970    & 30.5                          & -0.500           & -0.500           & 300                                           & $\pi/2$                                                               \\
\textbf{Injection 4}                     & 73.2                  & 18.3                  & 0.250    & 30.5                          & -0.600           & 0.300            & 300                                           & $\pi/2$                                                               \\
\textbf{Injection 5}                     & 20.3                  & 19.7                  & 0.970    & 17.4                          & 0.000            & 0.000            & 400                                           & $\pi/2$                                                               \\
\textbf{Injection 6}                     & 71.5                  & 35.8                  & 0.500    & 43.5                          & 0.200            & 0.200            & 200                                           & $\pi/2$                                                               \\
\textbf{Injection 7}                     & 50.8                  & 49.2                  & 0.970    & 43.5                          & 0.800            & 0.800            & 300                                           & $\pi/2$                                                               \\
\multicolumn{1}{c}{\textbf{Injection 8}} & 50.1                  & 25.1                  & 0.500    & 30.5                          & -0.600           & -0.600           & 300                                           & $\pi/2$                                                               \\ \bottomrule
\end{tabular}
}
\caption{List of injected parameters: individual masses, mass ratio, chirp mass, individual spin $z$ components, luminosity distance, and inclination angle for the eight zero-noise injections in LIGO A$^{\#}$ sensitivity.}
\label{tab:injections}
\end{table}

In all cases except one, we set the inclination angle to $\pi/2$ radians, as this maximizes the amplitude of the $(2,0)$ mode and thereby enhances its impact on parameter recovery. To assess the relevance of this parameter, we compare Injection 1 and Injection 2, which have the same parameters except for the inclination angle, which we vary from $\theta_{\text{JN}}=\pi/2$ radians (edge-on) to $\theta_{\text{JN}}=0$ radians (face-on). In Sec.~\ref{sec:inclination} we present the comparison between these two cases.

The remaining injected parameters, not listed in the table, are common across all injections: declination $\delta = -0.492$ radians, right ascension $\alpha = 2.77$ radians, polarization angle $\psi = 2.71$ radians, phase $\phi = 5.43$ radians, and a trigger time corresponding to the first gravitational-wave detection, GW150914.

Tab.~\ref{tab:runtimes} summarizes the network SNRs of the injected waveforms and the runtimes of the injections with the two versions of the model implemented in {\tt LALSuite}. In all runs, we choose the number of live points to be {\tt nlive=500} and the number of accepted steps to be {\tt naccept=15} to reduce the computational cost since we checked that the posteriors are sufficiently well sampled, and increasing them to {\tt nlive=1000} and {\tt naccept=60} led to equivalent results. We set the reference frequency to be the same as the initial frequency, $f_{\text{ref}}=f_{\text{min}}=10$ Hz, with a sampling frequency of 4096 Hz and starting the likelihood integration at 20 Hz in order to have the modes up to $m=4$ in band.

Injections 1 and 2 are the least computationally demanding. We observe that, as expected, lowering the total mass of the system results in a longer runtime, as the waveforms become longer. This is the case of Injection 5, which is the one with the lowest total mass. Moreover, increasing the magnitude of the spins and going towards more unequal masses increase the complexity of the waveforms and also leads to larger runtimes. In all the cases, we find that the addition of the $(2,0)$ mode in the analysis does not cause a significant computational overhead, as the difference in runtimes ranges from a few minutes to $\sim4$ hours in the most expensive case.

\begin{table}[h!]
\centering
\resizebox{\columnwidth}{!}{ 
\begin{tabular}{cccc}
\toprule
                     & \textbf{\begin{tabular}[c]{@{}c@{}}Network SNR\\ injected WF\end{tabular}} & \textbf{\begin{tabular}[c]{@{}c@{}}Runtime\\ IMRPhenomTHM+(2,0)\end{tabular}} & \textbf{\begin{tabular}[c]{@{}c@{}}Runtime\\ IMRPhenomTHM\end{tabular}} \\ \hline
\textbf{Injection 1} & 456                                                                        & 14h 52min                                                                     & 14h 49min                                                               \\
\textbf{Injection 2} & 930                                                                        & 13h 35min                                                                     & 12h 43min                                                               \\
\textbf{Injection 3} & 192                                                                        & 17h 56min                                                                     & 17h 19min                                                               \\
\textbf{Injection 4} & 184                                                                        & 17h 21min                                                                     & 17h 6min                                                                \\
\textbf{Injection 5} & 107                                                                        & 1d 21h 18min                                                                  &   1d 17h 25min                                                             \\
\textbf{Injection 6} & 435                                                                        & 15h 41min                                                                     & 15h 30min                                                               \\
\textbf{Injection 7} & 370                                                                        & 15h 38min                                                                     & 15h 13min                                                               \\
\textbf{Injection 8} & 174                                                                        & 21h 21min                                                                     &   18h 36min                                                                   \\ \bottomrule
\end{tabular}
}
\caption{Details about the set of injections performed for A$^{\#}$ sensitivity using 128 cores in 1 node with the {\tt LALSuite} implementation of the model. The parameters of each of the injections can be checked in Tab.~\ref{tab:injections}. The first column indicates the network (H1 and L1) SNR of the injected waveforms, the second column stands for the runtime of the recovery with the model including the $(2,0)$ mode, and the third column stands for the runtime of the recovery using the model without this mode.}
\label{tab:runtimes}
\end{table}
In the figures presented in this and the following section, we display the posterior distributions using corner plots. The one-dimensional marginal distributions of each parameter are shown along the top and right edges of the two-dimensional plots. In these plots, the black solid lines indicate the injected parameter values, while the dashed lines represent the 68\% ($1\sigma$) credible intervals of the posteriors. In the plot titles, the black value corresponds to the injected parameter, and each colored value denotes the median along with the 16th and 84th percentiles of the posterior distribution, using colors that match the distributions in the plot. In the two-dimensional distributions, the injected values are marked by a black star, and the medians of the posterior distributions are indicated with colored stars corresponding to the models used for recovery, as specified in the legends.

In Figs.~\ref{fig:posteriors_mass}-\ref{fig:distance_inclination} (\ref{fig:posteriors_CE}, \ref{fig:posteriors_ET}) the green (turquoise) distributions correspond to the recovery with the model including the $(2,0)$ mode, while the orange (pink) show the recovery with the model without including this mode. In Fig.~\ref{fig:posteriors_faceon_edgeon} we compare the recovery with the $(2,0)$ in both cases, but in blue the case of the edge-on injection and in red the one of the face-on injection.

In these 2D plots, for clarity, we only show the 90\% contour, while in the full corner plots, presented in Appendix \ref{app:cornerplots} and \ref{app:cornerplots_ET_CE}, the 68\%, 90\%, and 99\% credible regions are shown.

In the following subsections, we present and analyze the posterior distributions of the mass and spin parameters and investigate the impact that adding the $(2,0)$ spherical harmonic mode has on the distance-inclination degeneracy. Finally, we focus on the comparison between an edge-on and a face-on system to assess how the posteriors differ between these two orientations. 

\subsection{Mass parameters}
We begin by analyzing the recovery of the mass parameters. Fig.~\ref{fig:posteriors_mass} displays the joint posterior distributions for the mass ratio and the chirp mass for each of the eight injections. The posteriors of the individual component masses for each of the injections can be found in the complete corner plots presented in Appendix \ref{app:cornerplots}.

Overall, the mass ratio is well constrained across all injections and for both versions of the model. The maximum observed relative bias in the recovered mass ratio when neglecting the $(2,0)$ mode does not exceed 1\%. For Injection 3, a secondary mode appears in the posterior distribution of the chirp mass, which is slightly more pronounced when the $(2,0)$ mode is neglected, although it remains significantly smaller than the primary mode. As the binary's mass asymmetry increases, the chirp mass bias becomes more pronounced. This trend is most evident in Injection 4, representing the most asymmetric system considered ($q=0.25$), where the injected value of the chirp mass lies outside the 1$\sigma$ credible interval.

Focusing on Injection 2, which is the one corresponding to the face-on binary, we see that the distributions of both parameters are very similar in both models. However, if we compare this to Injection 1, which is the edge-on analog, we notice that the mass ratio distributions appear considerably different between the two injections. A more detailed comparison of these cases is presented in Sec.~\ref{sec:faceon_edgeon}.

\begin{widetext}
\begin{center}
\begin{figure}[htp]
\includegraphics[width=1\textwidth]{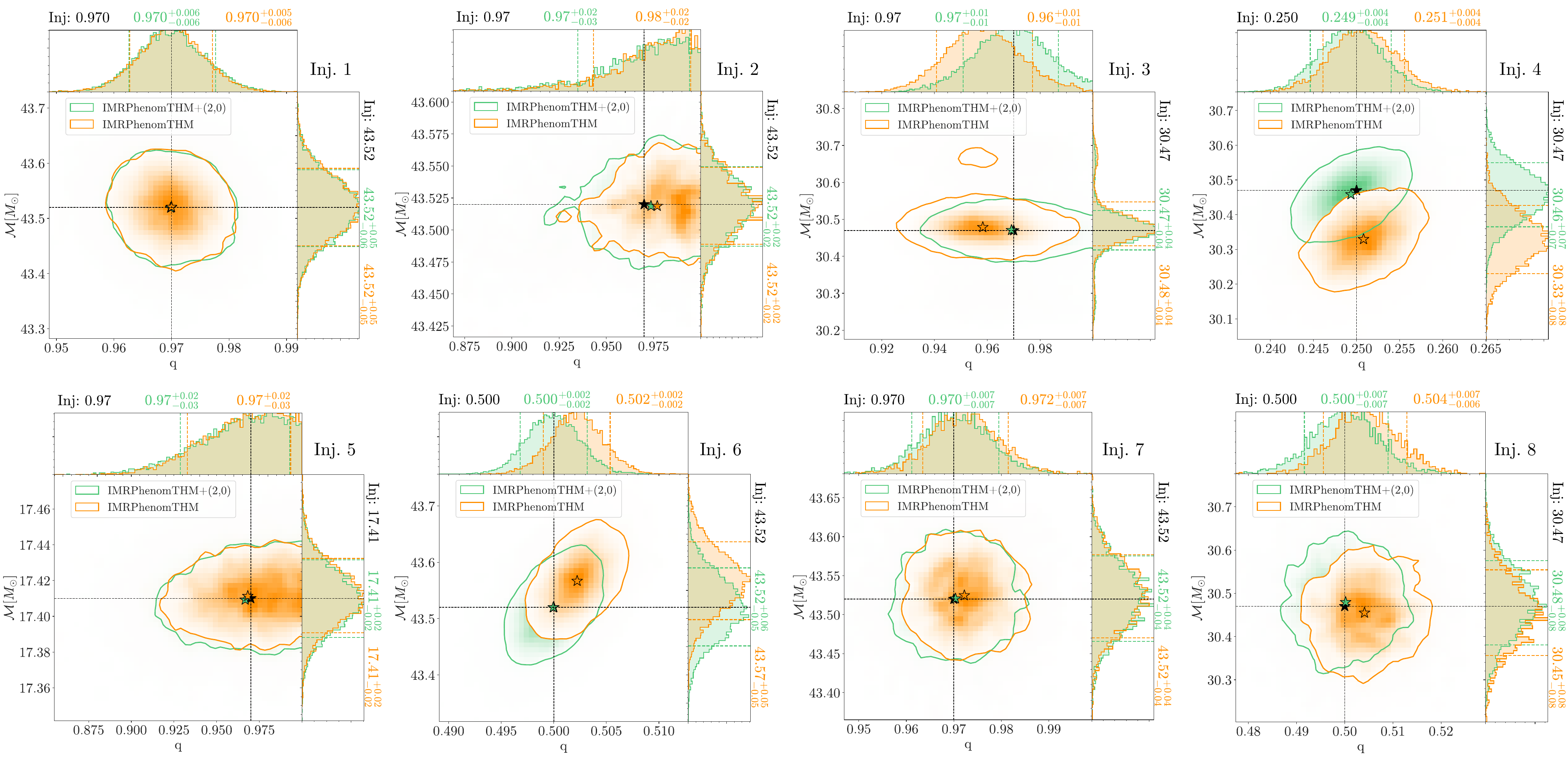}
    \caption{Joint posterior distributions of the mass ratio ($q$) and chirp mass ($\mathcal{M}$) for the eight injections in order from left to right and top to bottom. The black lines correspond to the injected parameter values.}
    \label{fig:posteriors_mass}
\end{figure}
\end{center}
\end{widetext}

\subsection{Spin parameters}
We now examine the recovery of the spin parameters. Fig.~\ref{fig:posteriors_spins} shows the joint posterior distributions of the dimensionless individual spin components for each of the eight injections. In the case of the nonspinning system (Injection 5), both models give posterior distributions consistent with the injected values, being able to perfectly recover the spin components when including the $(2,0)$ mode in the sampling.

In systems with moderate (Injections 1 and 6) and high (Injection 7) positive spins in an edge-on orientation, slight biases emerge when the $(2,0)$ mode is neglected. These biases place the injected values just outside or at the boundary of the 1$\sigma$ credible intervals. Comparison between Injection 1 and 2 is discussed in detail in Sec.~\ref{sec:faceon_edgeon}.

For the case of unequal spins (Injection 4), it can be noted that the posteriors, including the $(2,0)$ mode, are slightly biased. However, when neglecting it, the recovery of the secondary spin component is notably worsened. This injection has a mass ratio $q=0.25$, which is the most asymmetric system that we consider. Since this adds complexity to the waveform, recovery is not expected to be as good as in more equal-mass systems.

The most significant biases due to the omission of the $(2,0)$ mode occur for Injection 3 and Injection 8, which correspond to the injections with negative spin components in an edge-on configuration. The recovery with the model excluding the $(2,0)$ mode produces a completely offset distribution, with a recovered value biased by approximately 38\% in the worst case (secondary spin component in Injection 8). By checking Fig.~\ref{fig:amplitudes}, we see that the waveforms of this mode for the case of almost equal-mass and negative spins correspond to the ones in which the oscillatory contribution is most relevant. Therefore, this can be the reason why neglecting the mode in the analysis gives rise to more important biases. See Fig. 6 from \cite{PhysRevD.110.084074} to visually check the phenomenology of the waveforms for different combinations of mass ratio and spin components, especially the top left panel, which corresponds to the waveform with negative spin components.

The complete corner plots in Appendix~\ref{app:cornerplots} allow for a detailed examination of the mass parameter estimation, revealing that the observed spin biases are also present in the individual masses. This outcome is consistent with the fact that mass and spin parameters are coupled in their impact on the waveform morphology. For instance, a waveform from a higher mass ratio system may resemble that of a lower mass ratio system with higher spin magnitudes, which explains the simultaneous appearance of biases in both parameter sets.

\begin{widetext}
\begin{center}
\begin{figure}[htp]
\includegraphics[width=1\textwidth]{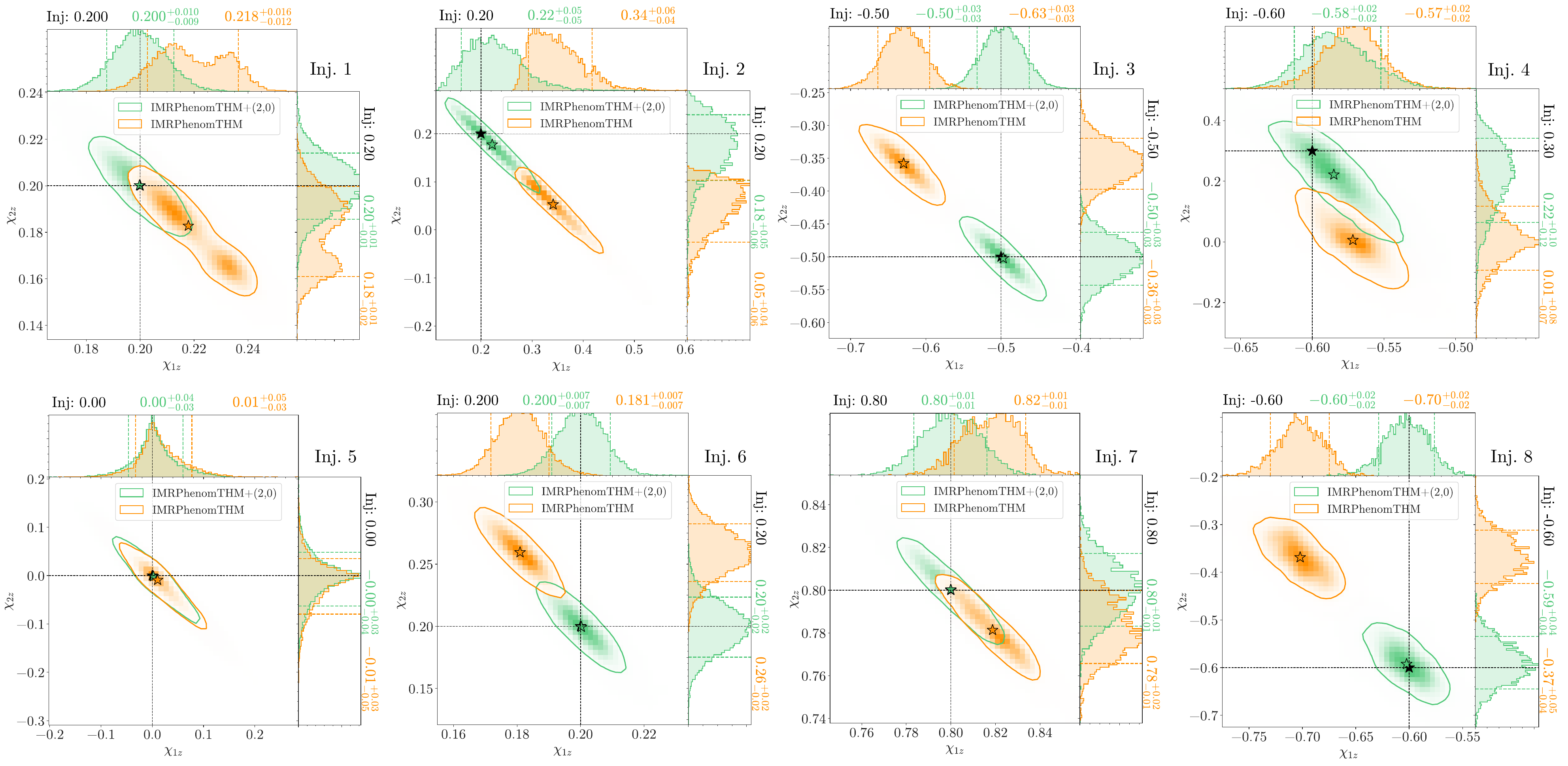}
    \caption{Joint posterior distributions of the individual spin $z$ components ($\chi_{1z}$ and $\chi_{2z}$) for the eight injections in order from left to right and top to bottom. The black lines correspond to the injected parameter values.}
    \label{fig:posteriors_spins}
\end{figure}
\end{center}
\end{widetext}

\subsection{Distance-inclination degeneracy}
The improvement of waveform models through the addition of physical effects results in a gain in the accuracy of the estimation of the source parameters, as we have seen. Moreover, it is useful to break the degeneracies present between some of the parameters, which is the case of distance-inclination degeneracy. Different distances scale the amplitude of the gravitational waves, becoming larger as the systems get closer. In a similar way, when just considering the dominant $(2,\pm2)$ modes, the factor given by the $s=-2$ spin-weighted spherical harmonic also acts as an overall scaling in the waveform, which is then degenerate with the distance. Since the dependence on the inclination is different for each of the modes, when including higher modes in the waveform, this degeneracy is broken. In this context, the case of the $(2,0)$ mode is particularly useful since its dependence on the inclination is completely opposed to the one of the $(2,\pm2)$. The spin-weighted spherical harmonic of the $(2,0)$ mode is given by
\label{sec:inclination}
\begin{equation}
    ^{-2}Y_{2,0}(\theta)=\frac{1}{4}\sqrt{\frac{15}{2\pi}}\sin^2\theta,
\end{equation}
from where we see that it is maximum for edge-on systems ($\theta=\pi/2$ radians) and vanishes for face-on ($\theta=0$ radians), in contrast to the main $(2,\pm2)$ modes, whose dependence on the inclination angle is given by
\begin{equation}
    ^{-2}Y_{2,\pm2}(\theta, \phi)=\sqrt{\frac{5}{64\pi}}(1\pm\cos\theta)^2e^{\pm2i\phi},
\end{equation}
which is maximized for face-on and face-off systems.

\begin{center}
\begin{figure}[htp]
\includegraphics[width=1\columnwidth]{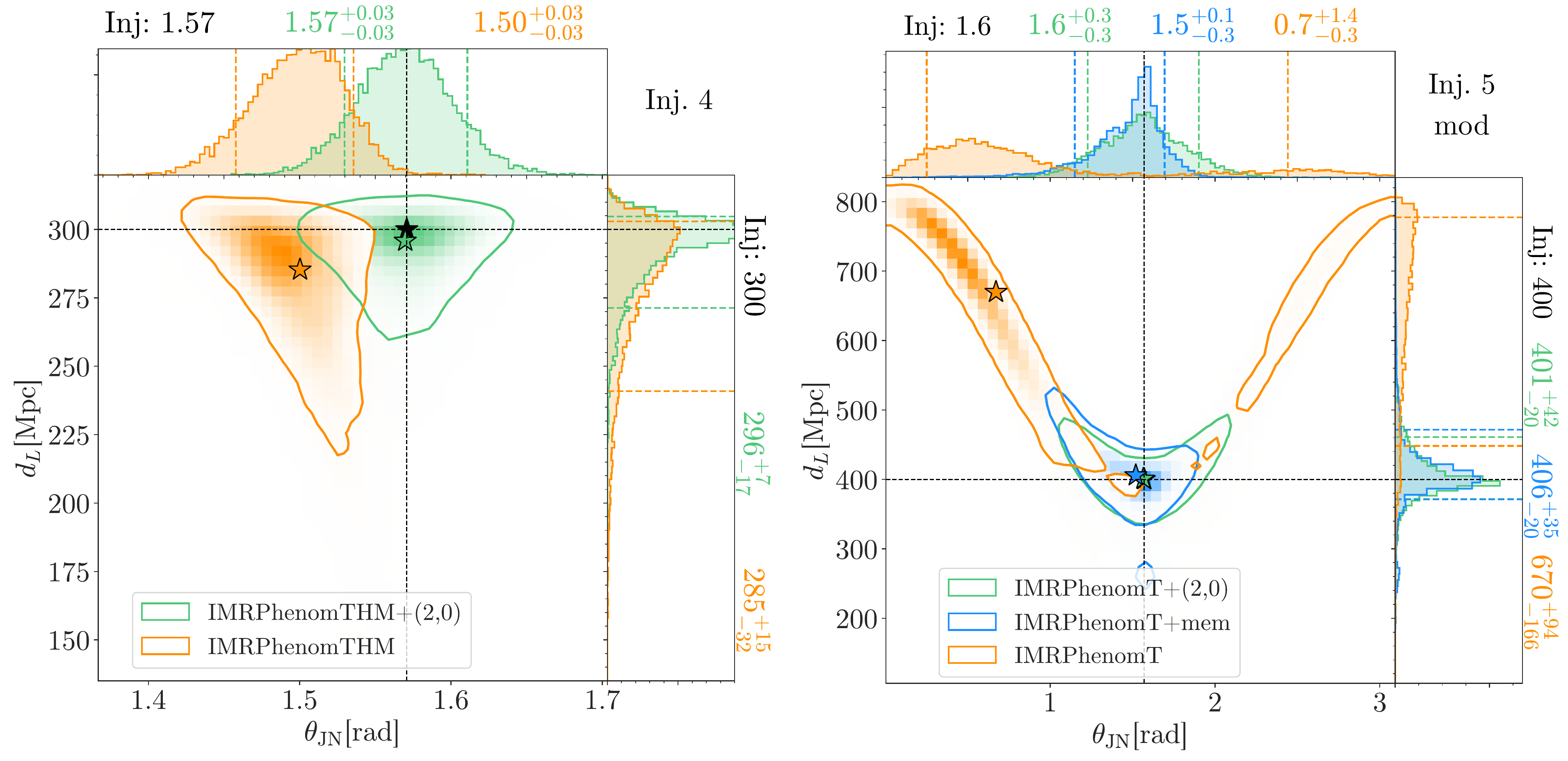}
    \caption{Joint posterior distributions of the luminosity distance ($d_L$) and inclination ($\theta_{\text{JN}}$). Left panel corresponds to Injection 4. Right panel compares the recovery with only the dominant $(2,\pm2)$ mode, adding the memory (result taken from \cite{Xu:2024ybt}), and adding the full $(2,0)$ mode when the higher modes are neglected both in the injection and the recovery. This corresponds to Injection 5 with the modified sky location and trigger time. The black lines correspond to the injected parameter values.}
    \label{fig:distance_inclination}
\end{figure}
\end{center}

We first evaluate the impact of including the $(2,0)$ mode on the recovery of the luminosity distance and inclination for the set of eight injections that have been performed using {\tt IMRPhenomTHM}. Our results (which can be checked in Appendix~\ref{app:cornerplots}) indicate that, in general, the inclusion of this mode does not produce a significant improvement in the recovery of these parameters. This outcome is consistent with expectations, as the higher-order modes already present in {\tt IMRPhenomTHM} provide strong constraints on distance and inclination, making the additional contribution of the $(2,0)$ mode negligible. The posterior distributions of these parameters remain almost identical for both model versions across most injections, with one notable exception: Injection 4 (see left panel of Fig.~\ref{fig:distance_inclination}). This corresponds to a binary system with a mass ratio of $q=0.25$, the model that neglects the $(2,0)$ mode exhibits a pronounced bias in the recovery of both the inclination angle and the luminosity distance. As the binary becomes more asymmetric, the oscillatory component of the $(2,0)$ mode is enhanced. Therefore, the biases observed in this injection are mainly caused by the absence of this contribution in the recovery, which turns out to be relevant, especially for the determination of the inclination.

Moving now to a different analysis of the distance-inclination degeneracy, we explore the effect of the $(2,0)$ mode when higher-order modes are absent. We perform an additional injection using the {\tt IMRPhenomT+(2,0)} model, which only includes the dominant $(2,\pm2)$ modes plus the $(2,0)$ and we recover the parameters with this mode content and only with the dominant modes to investigate the improvement of just adding the $(2,0)$ mode.

With this configuration, we aim to compare our results, where we consider the full $(2,0)$ mode (including both the memory and oscillatory contributions), with the results of the previous work presented in \cite{Xu:2024ybt}, where only the memory in this mode was considered. To do this, we set the same parameters, which correspond to those of Injection 5 with modifications to $\delta=0.811$ radians, $\alpha=1.96$ radians, and $t_c=1368057618$ s. The corresponding posterior distributions are shown in the right panel of  Fig.~\ref{fig:distance_inclination}.

The results reveal that, when using only the dominant $(2,\pm2)$ modes without the $(2,0)$ mode, the posteriors for distance and inclination are poorly constrained. The inclination posterior spans the full prior range, with the distribution peak far from the injected value. The distance posterior is similarly broad, significantly overestimating the injected value. These findings are consistent with the expected degeneracy between distance and inclination when only the dominant modes are included. In contrast, when adding the memory and the full $(2,0)$ mode, both parameters are much better determined and with considerably lower uncertainty. Although for these parameters (almost equal-mass, nonspinning), the memory contribution is more important than the oscillatory component of the mode, when considering the full mode, the recovered values lie closer to the injected ones. The findings using the full $(2,0)$ mode are consistent with those in \cite{Xu:2024ybt}. Based on the extended analysis presented in this work, which includes all subdominant harmonics in {\tt IMRPhenomTHM}, we find that the $(2,0)$ mode plays a crucial role in constraining the distance and inclination of the binary when only the dominant modes are present in the signal. However, once the full set of higher modes is included, the $(2,0)$ mode does not lead to a significant improvement in the recovery of these parameters in most of the cases that have been studied.

\subsection{Comparison edge-on and face-on binaries}
\label{sec:faceon_edgeon}
\begin{figure}[htp]
\includegraphics[width=1\columnwidth]{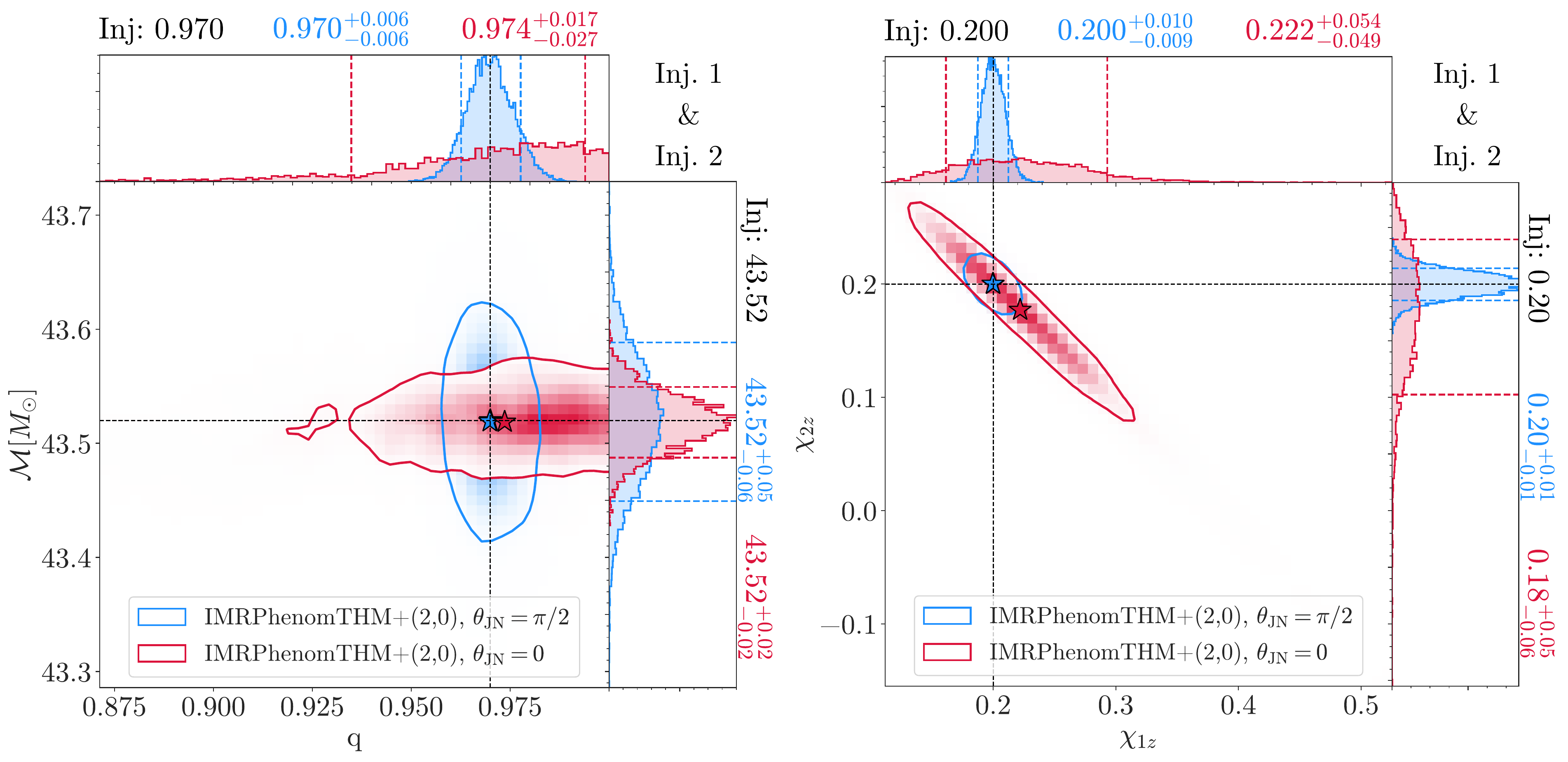}
    \caption{Joint posterior distributions of the mass ratio ($q$) and the chirp mass ($\mathcal{M}$) in the left panel and of the individual spin $z$ components ($\chi_{1z}$ and $\chi_{2z}$) in the right panel. Blue corresponds to the edge-on injection and red to the face-on injection. The black lines correspond to the injected parameter values.}
    \label{fig:posteriors_faceon_edgeon}
\end{figure}

We now focus on the comparison between the two injections with the same parameters, but different orientations: edge-on (Injection 1) and face-on (Injection 2), both including the $(2,0)$ mode in the recovery. As previously mentioned, this mode vanishes for face-on configurations, so it does not give a contribution to the injected waveform, but we want to inspect if it leads to an improvement in the recovered distributions when we include it in the analysis. Hence, in Fig.~\ref{fig:posteriors_faceon_edgeon} we do not compare the same injection with and without the $(2,0)$, but instead, the two different injections, both recovered, including this mode in the analysis, as stated in the legends. We show the joint distributions of mass ratio and chirp mass in the left panel and of the individual spin components in the right panel.

The mass ratio and spin components are much better constrained in the edge-on injection, whereas the chirp mass is more constrained in the face-on injection. This behavior arises because, for face-on orientations, the contribution of higher-order modes vanishes at $\theta=0$ radians. As a result, the signal effectively consists only of the dominant $(2,\pm2)$ modes, which are sufficient to accurately constrain the chirp mass due to their clean phase evolution. Furthermore, the face-on configuration results in a significantly higher signal-to-noise ratio (SNR), which enhances the precision of chirp mass estimation. In contrast, the edge-on signal has a lower overall SNR but contains non-negligible contributions from higher-order modes, which provide additional structure in the waveform. These features help break parameter degeneracies and lead to improved constraints on the mass ratio and spin components, as evident in the posterior distributions.

Another aspect of interest is the comparison of the posterior distributions of the reference phase ($\phi$) and the polarization angle ($\psi$) for both orientations (Injections 1 and 2) in Figs. \ref{fig:inj1} and \ref{fig:inj2}. In the first case, both parameters are accurately recovered, whereas in the second case, the posteriors remain uniform, essentially reflecting the prior distributions - uniform from 0 to $2\pi$ for the phase and from 0 to $\pi$ for the polarization angle. This suggests that the sampling has not constrained either of these two parameters in the case of Injection 2. This phenomenology was expected in the face-on scenario (as the higher modes vanish), due to the existing degeneracy between the reference phase and the polarization angle in waveforms that only contain the dominant mode, since the reference phase appears in the factor $e^{im\phi}$ contained in the spherical harmonics $^{-2}Y_{\ell,m}(\theta,\phi)$, which in this case has the same effect in the waveform as the polarization angle which appears as an overall factor $e^{i2\psi}$, causing the degeneracy between both parameters. The degeneracy is broken when subdominant harmonics are included in the waveform since the reference phase factor presents the dependence on $m$, which then affects each mode differently. Hence, this is the difference we can appreciate in the posteriors: for the edge-on inclination, the presence of higher modes breaks the degeneracy, whereas for the face-on orientation, no information can be retrieved for these two parameters.

\section{Injections combining LIGO A$^{\#}$ and 3G ground-based detectors}
\label{sec:3G}
We now turn to the case of combining the LIGO detectors (L1 and H1) with A$^{\#}$ sensitivity and 3G ground-based detectors: ET and CE, for which we show the sensitivity curves in Fig.~\ref{fig:PSDs}, which are taken from \cite{Sensitivitycurves}. Regarding parameter estimation studies for ET in the literature, there is a recent work \cite{mascioli2025tamingsystematicsdistanceinclination} that investigates the systematics in distance and inclination measurements for the different possible designs of this detector and the impact of higher modes when trying to break the degeneracy between these parameters. However, to date, there has been no investigation of parameter estimation that includes the $(2,0)$ spherical harmonic mode or the displacement memory effect for either ET or CE. We follow the same procedure as for the previous injections: we inject a signal with {\tt IMRPhenomTHM+(2,0)} and we recover the parameters with both {\tt IMRPhenomTHM+(2,0)} and {\tt IMRPhenomTHM}.

In our analysis, for ET, we use the equilateral triangle configuration (ET-1, ET-2, and ET-3 constituting ET-$\Delta$) of 10 km-long arms located at the Virgo site. For CE, we consider a single L-shaped detector with 40 km-long arms, located at the Hanford site, both as implemented in Bilby \cite{bilby_paper}. We consider two different networks: the first one consisting of CE together with L1, and the second one consisting of ET, L1, and H1. With these detector network configurations, we study the effect of replacing H1 by a more sensitive detector and the effect of adding a third detector (more sensitive at lower frequencies) to the LIGO network. We perform two injections with the set of parameters listed in Tab.~\ref{tab:injections_3G}. Injection 9 is analyzed using both detector networks, while Injection 10 is studied only for the ET+L1+H1 network.

\begin{table}[h!]
\centering
\resizebox{\columnwidth}{!}{ 
\begin{tabular}{ccccccccc}
\toprule
\multicolumn{1}{l}{}  & $\bm{m_1[M_{\odot}]}$ & $\bm{m_2[M_{\odot}]}$ & $\bm{q}$ & $\bm{\mathcal{M}[M_{\odot}]}$ & $\bm{\chi_{1z}}$ & $\bm{\chi_{2z}}$ & $\bm{d_L}$\textbf{[Mpc]} & $\bm{\theta_{\text{\textbf{JN}}}}$\textbf{[rad]} \\ \hline
\textbf{Injection 9}  & 50.8                    & 49.2                    & 0.970     & 43.5                            & 0.200             & 0.200             & 800                                           & $\pi/2$                                                               \\
\textbf{Injection 10} & 102                   & 98.5                    & 0.970     & 87.0                            & 0.600             & 0.600             & 800                                           & $\pi/2$                                                               \\ \bottomrule
\end{tabular}
}
\caption{List of injected parameters: individual masses, mass ratio, chirp mass, individual spin $z$ components, luminosity distance, and inclination angle for the zero-noise injections in CE+L1 and ET+L1+H1 networks of detectors.}
\label{tab:injections_3G}
\end{table}

The rest of the injected parameters not included in the table are: declination $\delta=-0.492$ radians, right ascension $\alpha=2.77$ radians, polarization angle $\psi=2.71$ radians, phase $\phi=5.43$ radians, and the trigger time corresponding to the first detection (GW150914). In these injections, we decrease $f_{\text{min}}$ to 5 Hz as ET and CE are more sensitive at low frequencies (see Fig.~\ref{fig:PSDs} \cite{Sensitivitycurves}). We keep $f_{\text{ref}}=10$ Hz, the sampling frequency to 4096 Hz, and we start the likelihood integration at 10 Hz, so that, again, the modes up to $m=4$ are in band. The sampler settings are the same as in the previous runs. In Tab.~\ref{tab:rutimes3G}, we show the network SNR and runtimes (when including and neglecting the $(2,0)$ mode in the recovery) for the injections in both detector networks.

We use the same priors as for Injections 1-8, listed in Tab.~\ref{tab:priors}, except that we modify the priors for the individual and chirp masses in order to reduce the runtimes. We employ uniform priors but restrict them more closely around the injected values. Therefore, for Injection 9 we set: $m_1,m_2\in\{45,65\}M_{\odot}$ and $\mathcal{M}\in\{35,55\}M_{\odot}$; and for Injection 10: $m_1,m_2\in\{90,110\}M_{\odot}$ and $\mathcal{M}\in\{75,95\}M_{\odot}$. As already seen in the previous injection studies, the addition of the $(2,0)$ mode in the recovery does not imply a significant increase in the runtimes.

\begin{table}[h!]
\centering
\resizebox{\columnwidth}{!}{
\begin{tabular}{ccccc}
\toprule
                      & \textbf{\begin{tabular}[c]{@{}c@{}}Detector\\ network\end{tabular}} & \textbf{\begin{tabular}[c]{@{}c@{}}Network SNR\\ injected WF\end{tabular}} & \textbf{\begin{tabular}[c]{@{}c@{}}Runtime\\ IMRPhenomTHM+(2,0)\end{tabular}} & \textbf{\begin{tabular}[c]{@{}c@{}}Runtime\\ IMRPhenomTHM\end{tabular}} \\ \hline
\textbf{Injection 9}  & CE+L1                                                               & 594                                                                        & 3d 15h 33min                                                                     &  3d 12h 1min                                                                 \\
\textbf{Injection 9}  & ET+L1+H1                                                            & 257                                                                        &   3d 6h 59min                                                                        & 3d 3h 7min                                                                        \\
\textbf{Injection 10} & ET+L1+H1                                                            & 484                                                                        & 23h 52min                                                                     & 23h 39min                                                               \\ \bottomrule
\end{tabular}
}
\caption{Details about the injections performed in the two network configurations that we consider, using 128 cores in 1 node with the {\tt LALSuite} implementation of the model. The first column indicates the detectors that compose the network, the second column stands for the SNR of the injected waveform, the third column states the runtime of the recovery with the model including the $(2,0)$ mode, and the fourth column for the runtime of the recovery using the model without this mode.}
\label{tab:rutimes3G}
\end{table}

In the following subsections, we present the results of the injections performed in the two networks, including 3G gravitational wave detectors.

\begin{widetext}
\begin{center}
\begin{figure}[htp]
\includegraphics[width=0.75\textwidth]{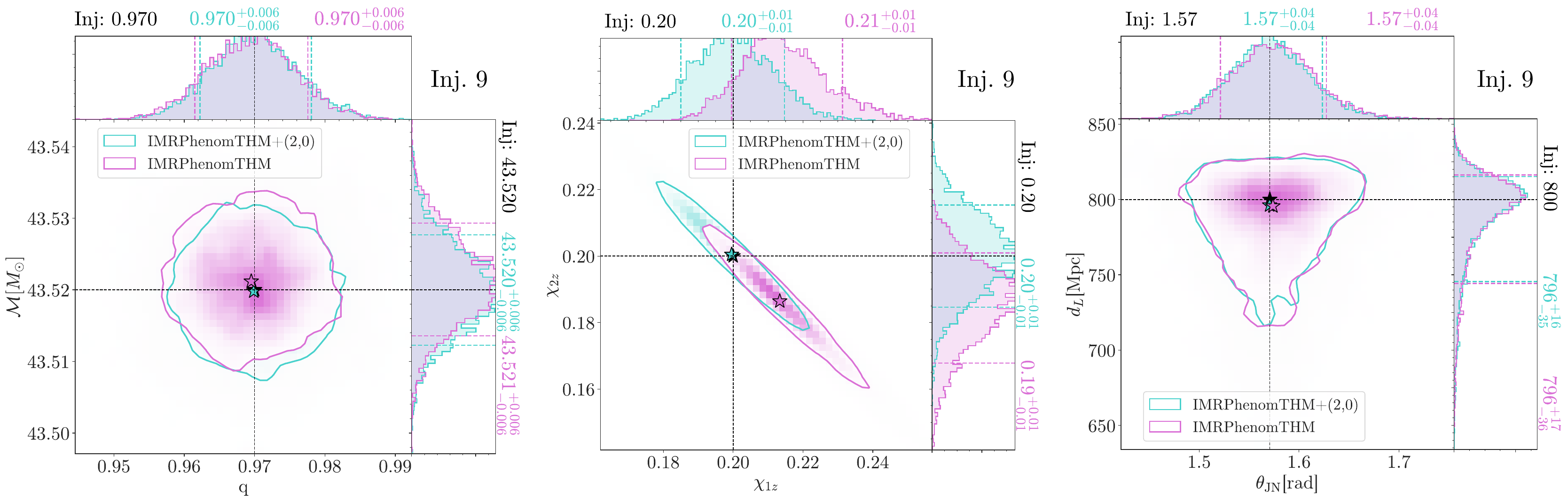}
    \caption{Joint posterior distributions for the detector network composed of CE and L1. The left column shows the mass ratio ($q$) and the chirp mass ($\mathcal{M}$); the center column shows the individual spin $z$ components ($\chi_{1z}$ and $\chi_{2z}$); and the right column shows the luminosity distance ($d_L$) and the inclination angle ($\theta_{\text{JN}}$). The injected parameters are the ones listed in Tab.~\ref{tab:injections_3G} as Injection 9. The black lines correspond to the injected parameter values.}
    \label{fig:posteriors_CE}
\end{figure}

\begin{figure}[htp]
\includegraphics[width=0.75\textwidth]{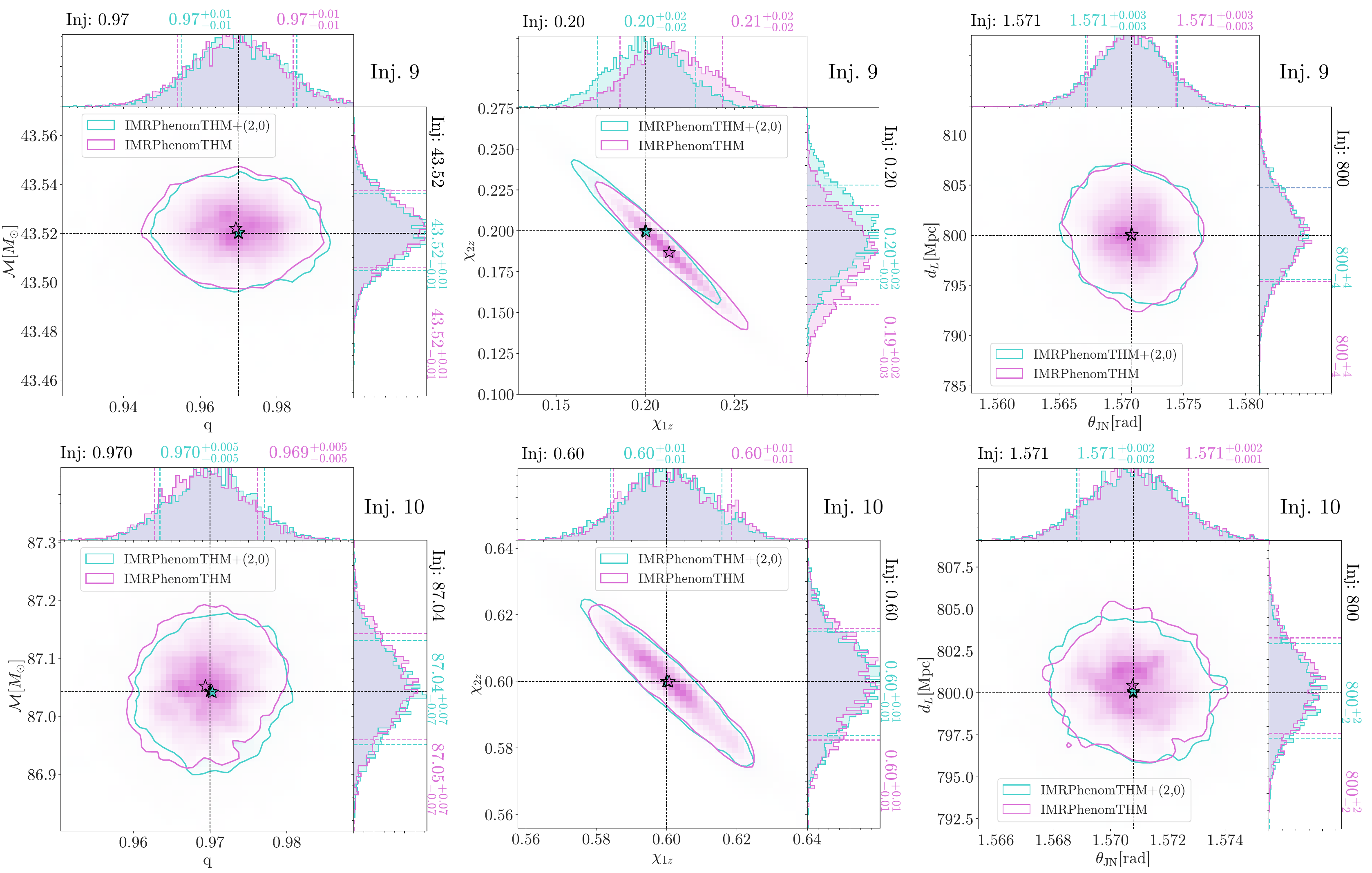}
    \caption{Joint posterior distributions for the detector network composed of ET, L1, and H1. The left column shows the mass ratio ($q$) and the chirp mass ($\mathcal{M}$); the center column shows the individual spin $z$ components ($\chi_{1z}$ and $\chi_{2z}$); and the right column shows the luminosity distance ($d_L$) and the inclination angle ($\theta_{\text{JN}}$). The top row corresponds to Injection 9, and the bottom row to Injection 10 in Tab.~\ref{tab:injections_3G}. The black lines indicate the injected parameter values.}
    \label{fig:posteriors_ET}
\end{figure}
\end{center}
\end{widetext}

\subsection{Network with LIGO Livingston and Cosmic Explorer}
First, we consider the detector network composed of CE and L1. We present in Fig.~\ref{fig:posteriors_CE} the joint 2D posterior distributions of mass, the ratio-chirp mass, individual $z$ spin components, and distance-inclination. Each of the plots shows the comparison when recovering with and without the $(2,0)$ mode in the waveforms as indicated in the legends. The full corner plots can be found in Fig.~\ref{fig:inj9CE} from Appendix \ref{app:cornerplots_ET_CE}. By checking the posterior distributions in the left plot, we note that the chirp mass and mass ratio are equally well determined, and the distance and inclination are well constrained using both models. Regarding the individual spin components in the middle plot, as seen in previous results, there is a bias in the recovery when neglecting the $(2,0)$ mode. The conclusions we can draw from these results are consistent with those found when analyzing the L1+H1 network.

More specifically, compared with Injection 1 - which has exactly the same parameters as this one except for the distance, which in this case is raised to 800 Mpc - we observe very similar behavior in all the parameters. Therefore, we do not perform further injection studies with this detector network as the results are qualitatively similar to those already obtained with the L1+H1 network.

\subsection{Network with both LIGO detectors and Einstein Telescope}
We now turn to the case where we consider a detector network composed of the two LIGO and ET observatories. We perform two different injections: Injection 9, which is the same as considered in the previous scenario with the CE+L1 network; and Injection 10, which has larger values of the total mass and spin magnitudes. Due to limited computational resources, we restrict the analysis to considerably high values of the total mass.

In Fig.~\ref{fig:posteriors_ET} we present the joint 2D posterior distributions of mass ratio-chirp mass, individual $z$ spin components, and distance-inclination of the two injections we consider in this detector network. For Injection 9, we retrieve posteriors very similar to the ones obtained using the L1+CE network for the intrinsic parameters, with a slight improvement in the spin components when including the $(2,0)$ mode in the analysis. This behavior is not reproduced in Injection 10 as the spin components are equally well recovered in both cases. Compared to the posterior distributions for the distance-inclination previously presented, in these cases, the shape of the distance posterior is symmetric, while in previous results, there was a tail at lower distances. This difference is thought to be due to the fact that this is a network composed of three detectors, which helps to determine the distance more accurately. This also happens for the parameters $\delta, \alpha$, and $\psi$, which are more constrained in these last two injections. The posteriors for these quantities and all the sample parameters can be found in the full corner plots in Figs.~\ref{fig:inj9ET} and \ref{fig:inj10ET} from Appendix~\ref{app:cornerplots_ET_CE}.

\section{Conclusions}
\label{sec:conclusions}
We have performed a PE injection-recovery campaign to investigate the biases introduced when the presence of the $(2,0)$ spherical harmonic mode is neglected. Our analysis involved three distinct detector networks. Initially, we performed eight injections across a range of parameters using the LIGO network with A$^{\#}$ sensitivity. We then extended the study to 3G networks, combining the LIGO detectors with the next-generation CE and ET detectors. Since we extend the frequency range by lowering the minimum frequency to 5 Hz, runs become more computationally expensive. Given that the results obtained with the 3G network did not significantly deviate from those of the L1+H1 network, we limited our analysis to one injection in the L1+CE network and two injections in the L1+H1+ET network.

The full corner plots as well as the relative biases in all the sampled parameters are presented in Appendices \ref{app:cornerplots} and \ref{app:cornerplots_ET_CE} for all the injections. In some of the parameters, the relative biases are slightly worse for the recovery, including the $(2,0)$ mode. However, in all these cases, the posterior distributions obtained with both versions of the model look practically identical, and the biases are so small that this suggests that the difference in the relative biases is caused by statistical variations during the sampling procedure.

Our findings indicate that, while the $(2,0)$ spherical harmonic mode is subdominant, its omission can introduce parameter biases at high SNRs. Unlike previous studies that focused only on the memory contribution in this mode, our results account for the complete mode, uncovering novel biases in the individual spin components, which can become significant for some combinations of parameters. In general, the improvement in the distance-inclination degeneracy is less significant compared to other higher-order modes, as they already constrain these parameters quite well, so that the contribution of the $(2,0)$ mode is not noticeable. We have performed a comparison injecting only the dominant $(2,\pm2)$ and $(2,0)$ modes and recovering with the $(2,\pm2)$, adding the full $(2,0)$ and adding only the memory contribution in the $(2,0)$, using the results from \cite{Xu:2024ybt}. In this situation, the particular dependence on the inclination of the $(2,0)$ becomes useful to break the distance-inclination degeneracy in the absence of other higher modes, obtaining consistent results with previous studies.

We expect that parameter biases become more pronounced at lower masses, though this investigation is limited by computational cost. Adding future detectors to the networks considerably increases the SNR of the signals, which helps the determination of the source parameters, supporting the strategy of combining data from current and next-generation detectors to enhance sensitivity. Given the size of the parameter space, fully exploring it using Bayesian parameter estimation injections becomes computationally prohibitive. Performing Fisher matrix studies across a broader parameter space would be a computationally much cheaper approach, which we leave as future work.

\section*{Acknowledgments}

The authors gratefully thank Maria de Lluc Planas for providing help in understanding the Fourier Transform of the time-domain waveforms within the Bilby code, Jorge Valencia for useful discussions and suggestions in the presentation of the results of this work and Cecilio García-Quirós for helpful comments on the manuscript. 

The authors thankfully acknowledge the computer resources at Picasso and the technical support provided by Barcelona Supercomputing Center (BSC) through grant No. AECT-2024-3-0019 from the Red Española Supercomputación (RES). The authors thank the Supercomputing and Bioinnovation Center (SCBI) of the University of Málaga for their provision of computational resources and technical support (\url{www.scbi.uma.es/site}). M.R.-S. is supported by the Spanish Ministry of Universities via an FPU doctoral grant (FPU21/05009). This work was supported by the Universitat de les Illes Balears (UIB); the Spanish Agencia Estatal de Investigación grants PID2022-138626NB-I00, RED2022-134204-E, RED2022-134411-T, funded by MICIU/AEI/10.13039/501100011033 and by the ESF+ and the ERDF/EU; and the Comunitat Autònoma de les Illes Balears through the Conselleria d'Educació i Universitats with funds from the European Union - NextGenerationEU/PRTR-C17.I1 (SINCO2022/6719) and from the European Union - European Regional Development Fund (ERDF) (SINCO2022/18146). S. B. received partial support from the French government under the France 2030 investment plan, as part of the Initiative d'Excellence d'Aix Marseille Université - A*MIDEX AMX-22-CEI-02.

\clearpage
\appendix
\begin{widetext}
\section{Full corner plots for the injections in LIGO A$^{\#}$ network}
\label{app:cornerplots}
In Figs.~\ref{fig:inj1}-\ref{fig:inj8}, we show the full corner plots of the posterior distributions of the sampled parameters for the injections performed with the two LIGO detectors at A$^{\#}$ sensitivity. We include the 2D joint posteriors of the parameters, and on the diagonal of the figures, we show the 1D marginal distributions. The green distributions correspond to the recovery with the model including the $(2,0)$ mode, while the orange show the recovery with the model neglecting this mode. The black lines and stars show the injected values of the parameters. The contours in the 2D plots represent the 68\%, 95\%, and 99\% credible intervals, while the vertical dashed lines in the 1D posteriors - as well as the subscripts and superscripts on the median shown in the plot titles - indicate the $1\sigma$ (68\%) credible intervals of the posterior distributions. The top right panels in the figures present the relative bias in percentage for each of the sampled parameters and each version of the model. Green dots stand for the model including the $(2,0)$ mode, while orange triangles represent the model neglecting this mode.

\begin{figure}[htp]
\includegraphics[width=1\textwidth]{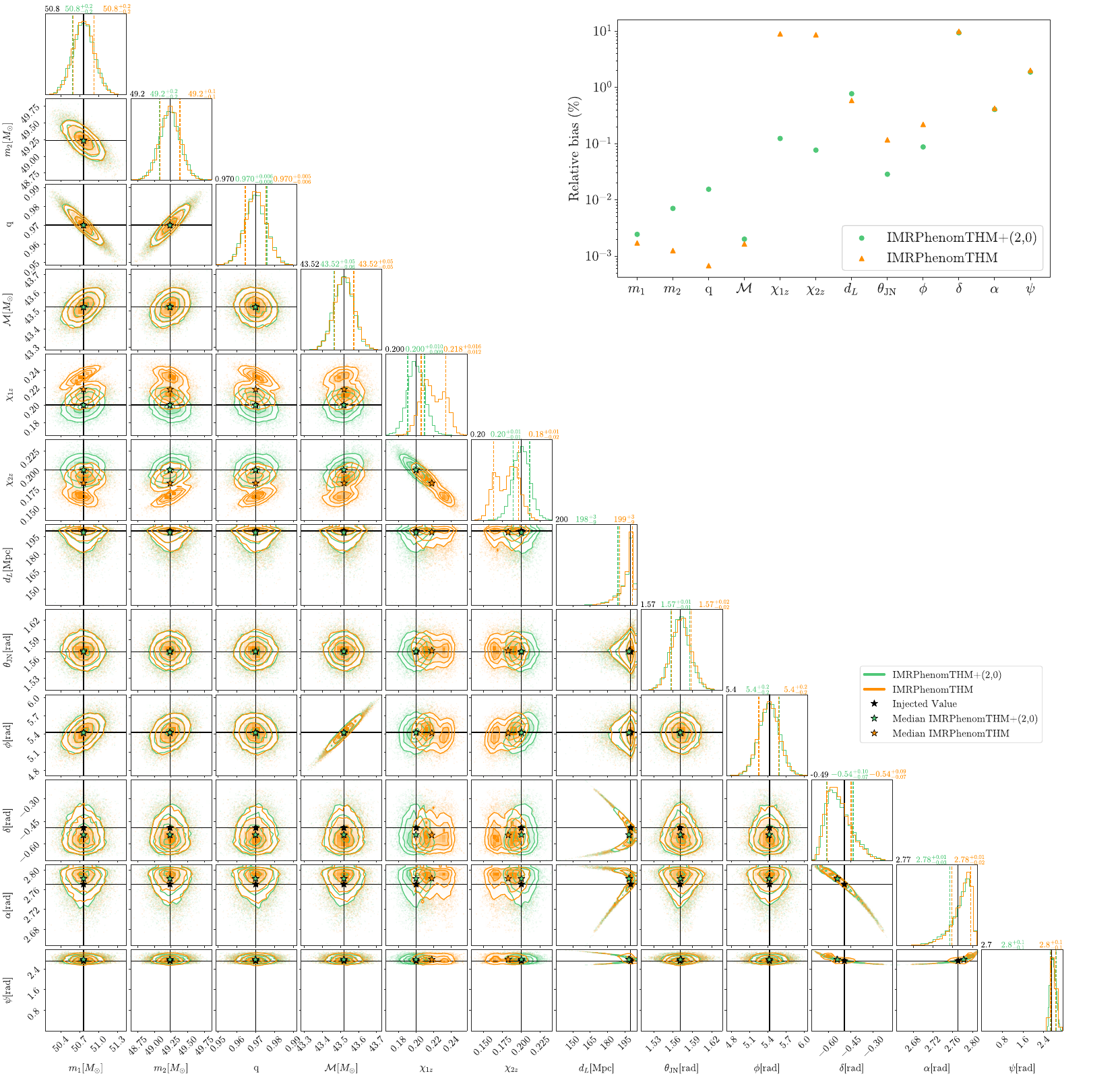}
    \caption{Full corner plot and relative bias for Injection 1 in the LIGO network.}
    \label{fig:inj1}
\end{figure}

\newpage
\thispagestyle{empty} 
\vspace*{\fill} 
\begin{figure}[htp]
\centering
\includegraphics[width=1\textwidth]{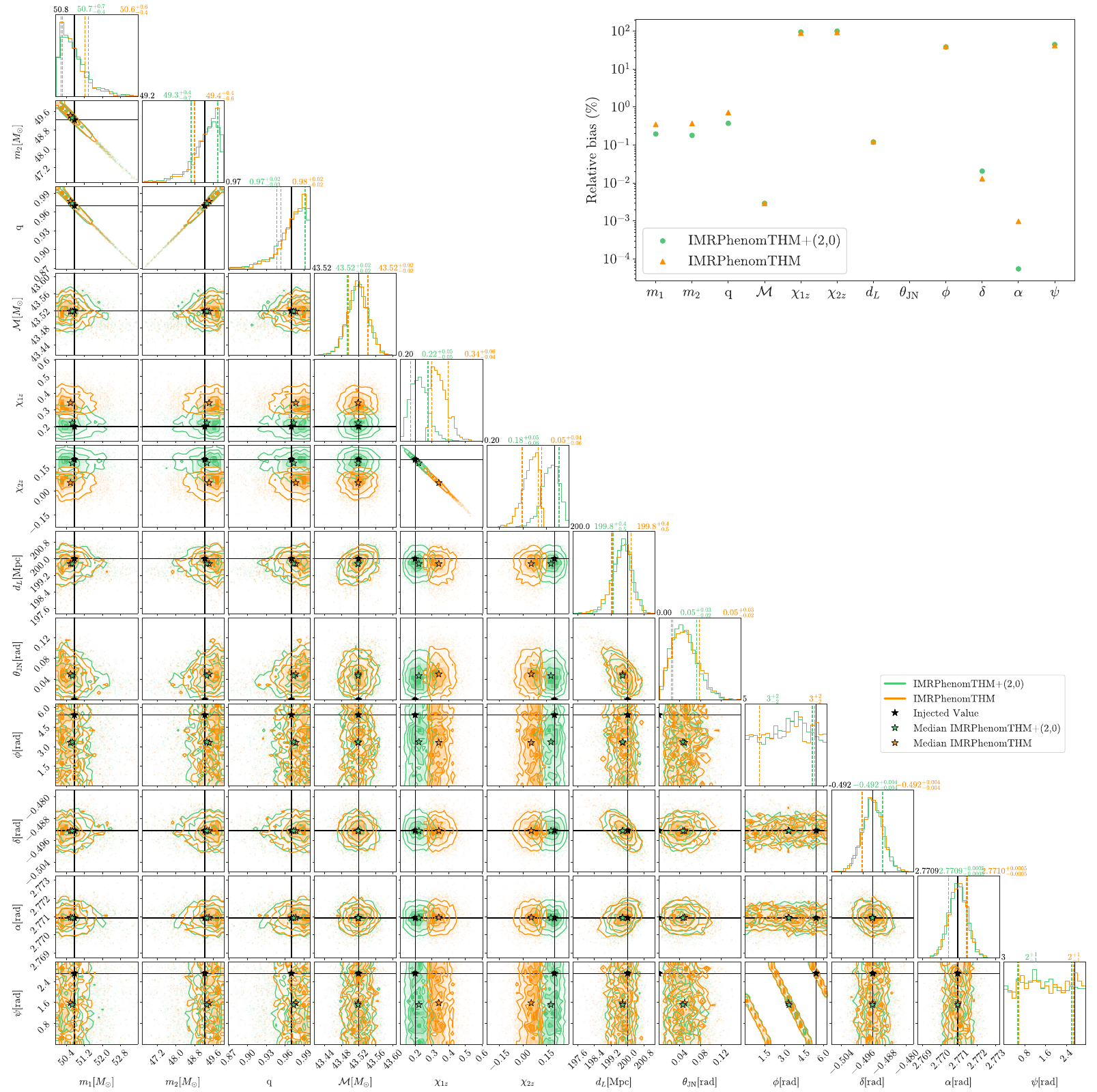}
\caption{Full corner plot and relative bias for Injection 2 in the LIGO network.}
\label{fig:inj2}
\end{figure}
\vspace*{\fill} 

\newpage
\thispagestyle{empty} 
\vspace*{\fill} 
\begin{figure}[htp]
\centering
\includegraphics[width=1\textwidth]{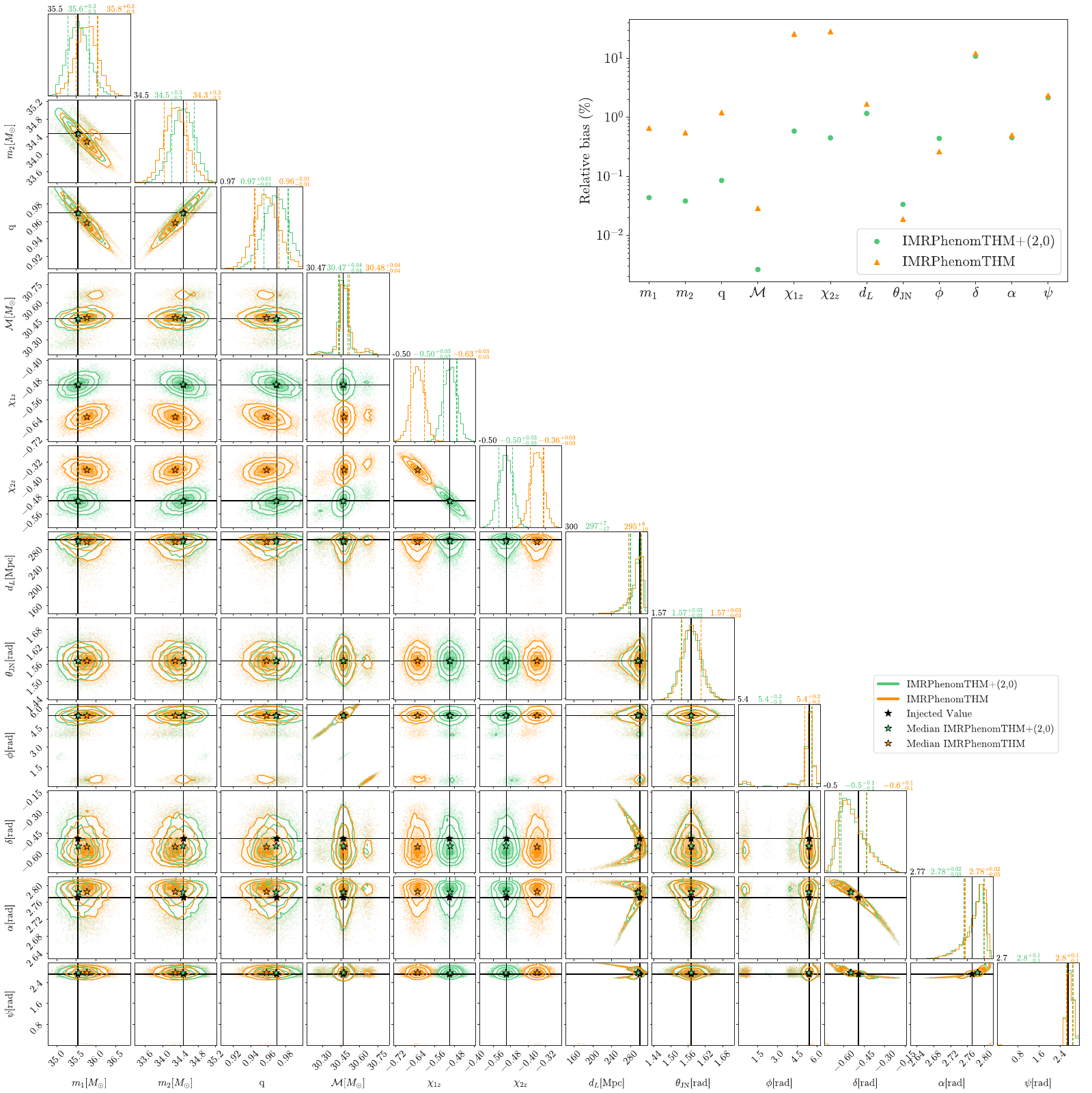}
\caption{Full corner plot and relative bias for Injection 3 in the LIGO network.}
\label{fig:inj3}
\end{figure}
\vspace*{\fill} 

\newpage
\thispagestyle{empty} 
\vspace*{\fill} 
\begin{figure}[htp]
\centering
\includegraphics[width=1\textwidth]{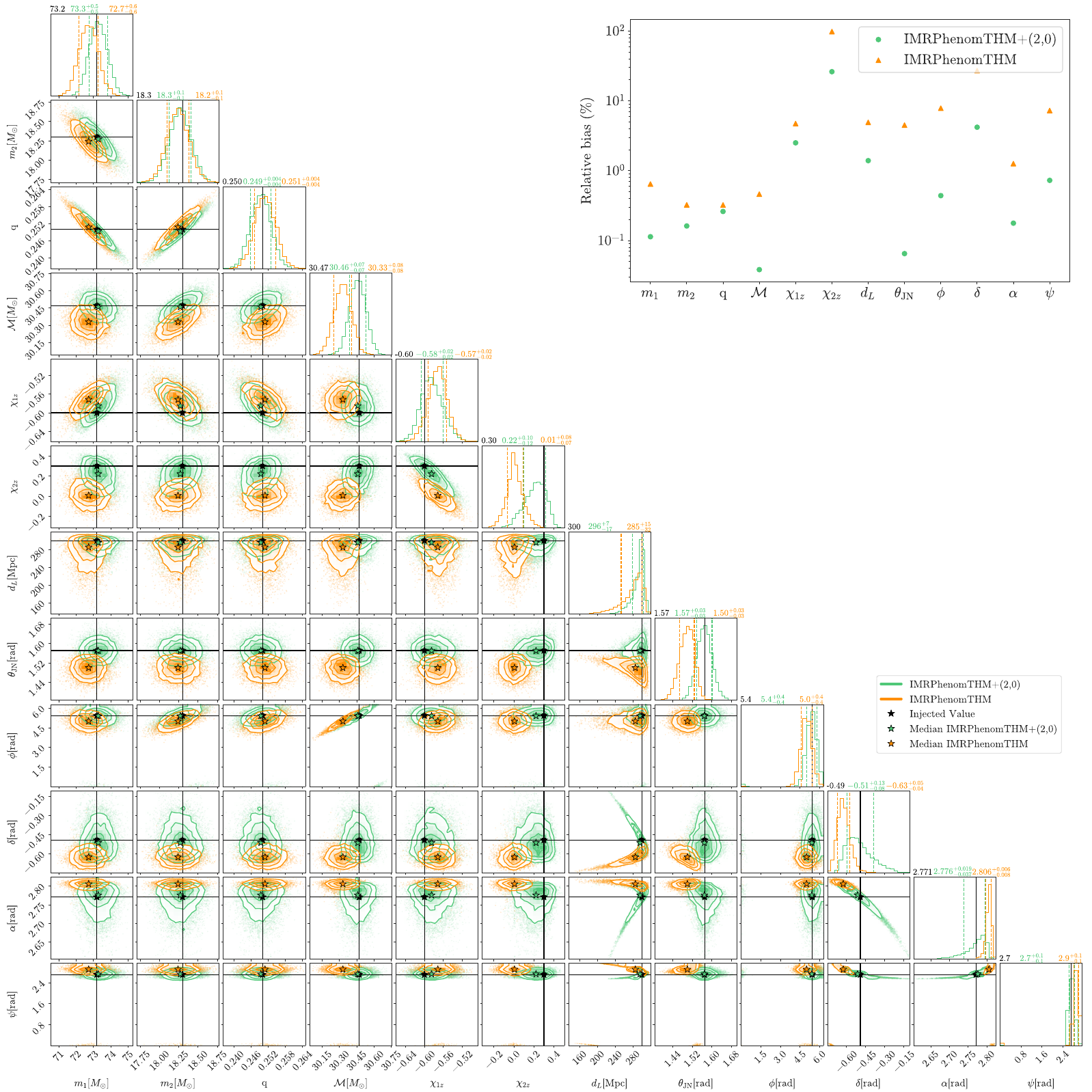}
\caption{Full corner plot and relative bias for Injection 4 in the LIGO network.}
\label{fig:inj4}
\end{figure}
\vspace*{\fill} 

\newpage
\thispagestyle{empty} 
\vspace*{\fill} 
\begin{figure}[htp]
\centering
\includegraphics[width=1\textwidth]{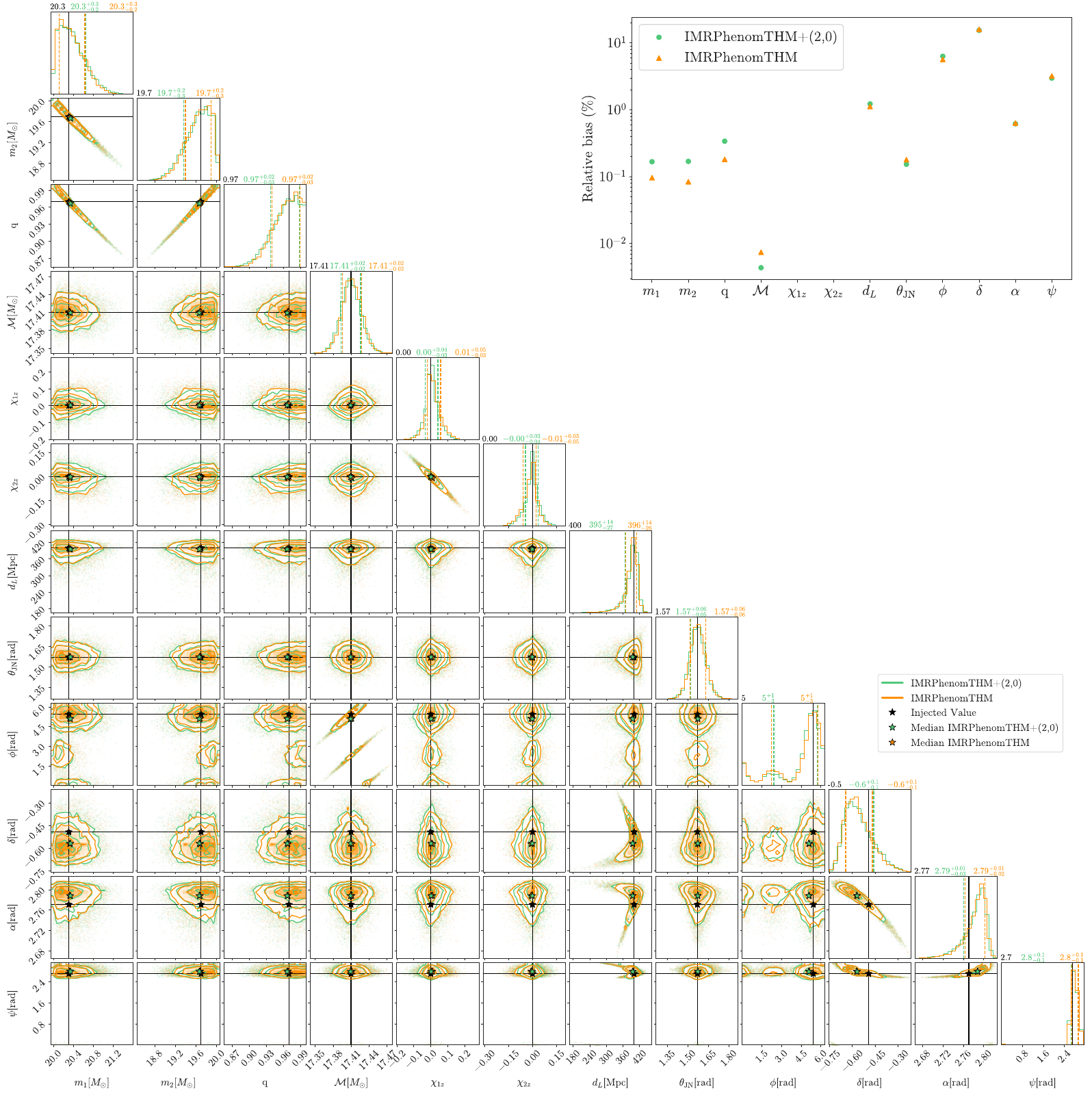}
\caption{Full corner plot and relative bias for Injection 5 in the LIGO network.}
\label{fig:inj5}
\end{figure}
\vspace*{\fill} 

\newpage
\thispagestyle{empty} 
\vspace*{\fill} 
\begin{figure}[htp]
\centering
\includegraphics[width=1\textwidth]{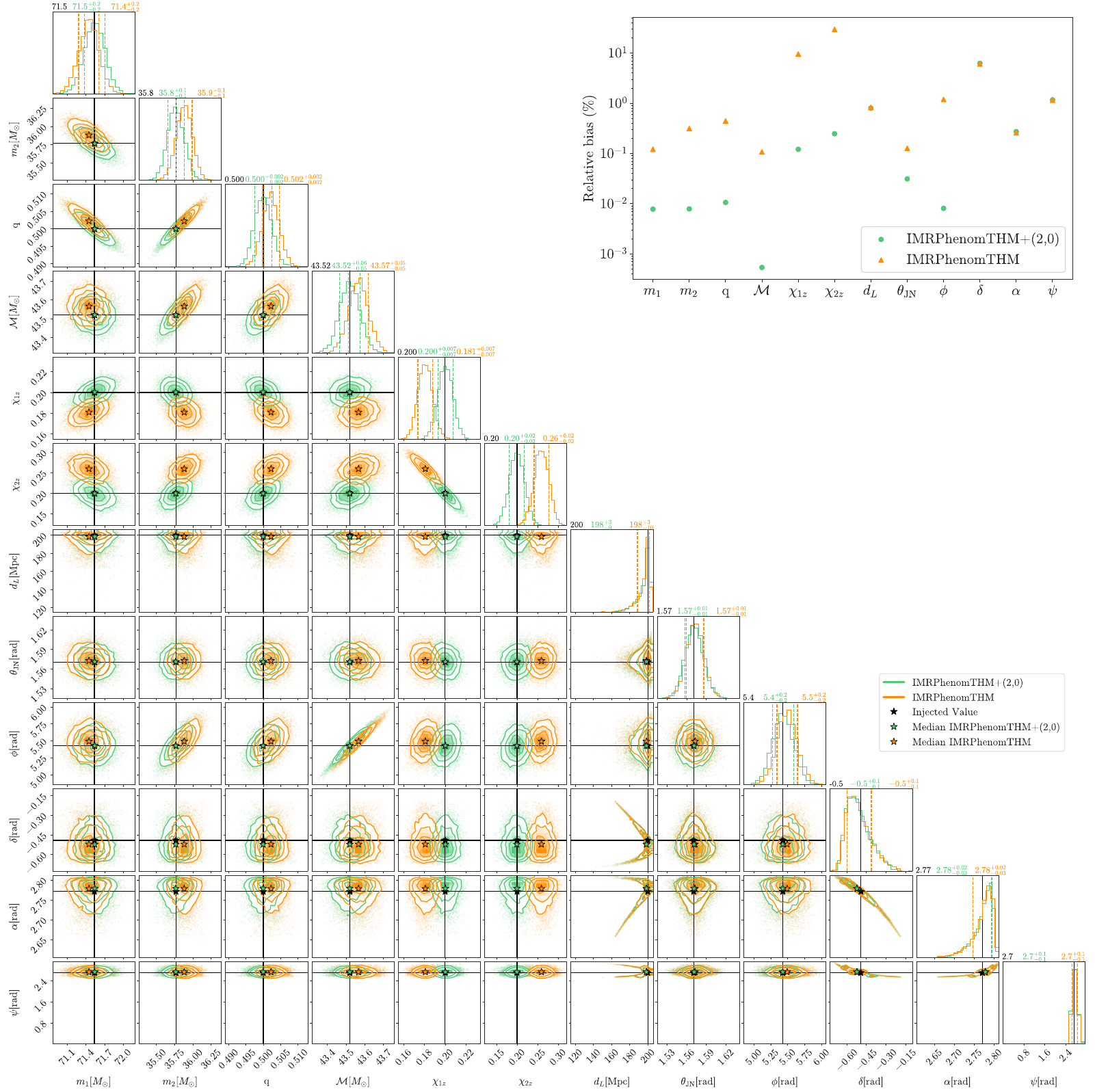}
\caption{Full corner plot and relative bias for Injection 6 in the LIGO network.}
\label{fig:inj6}
\end{figure}
\vspace*{\fill} 

\newpage
\thispagestyle{empty} 
\vspace*{\fill} 
\begin{figure}[htp]
\centering
\includegraphics[width=1\textwidth]{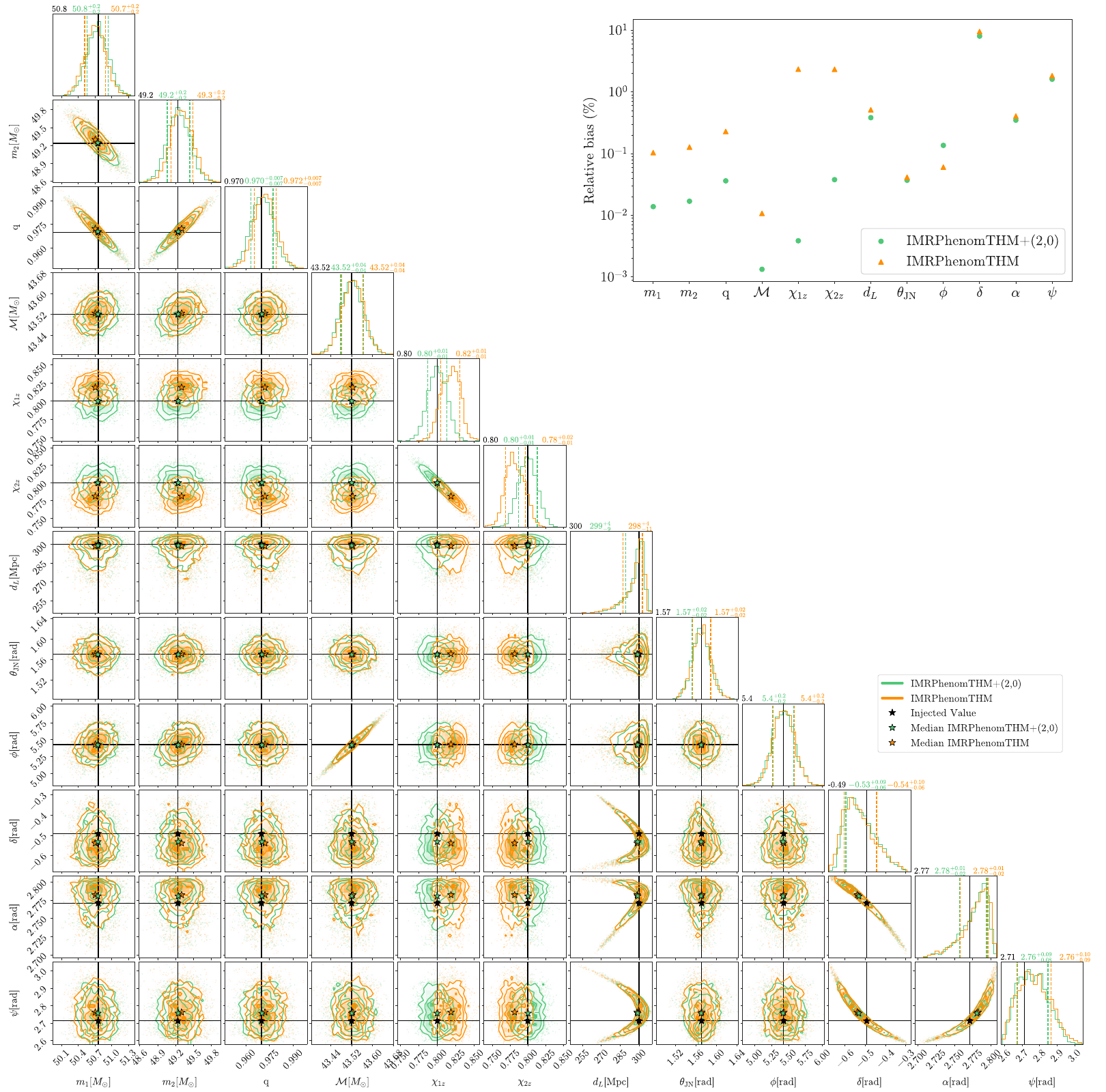}
\caption{Full corner plot and relative bias for Injection 7 in the LIGO network.}
\label{fig:inj7}
\end{figure}
\vspace*{\fill} 

\newpage
\thispagestyle{empty} 
\vspace*{\fill} 
\begin{figure}[htp]
\centering
\includegraphics[width=1\textwidth]{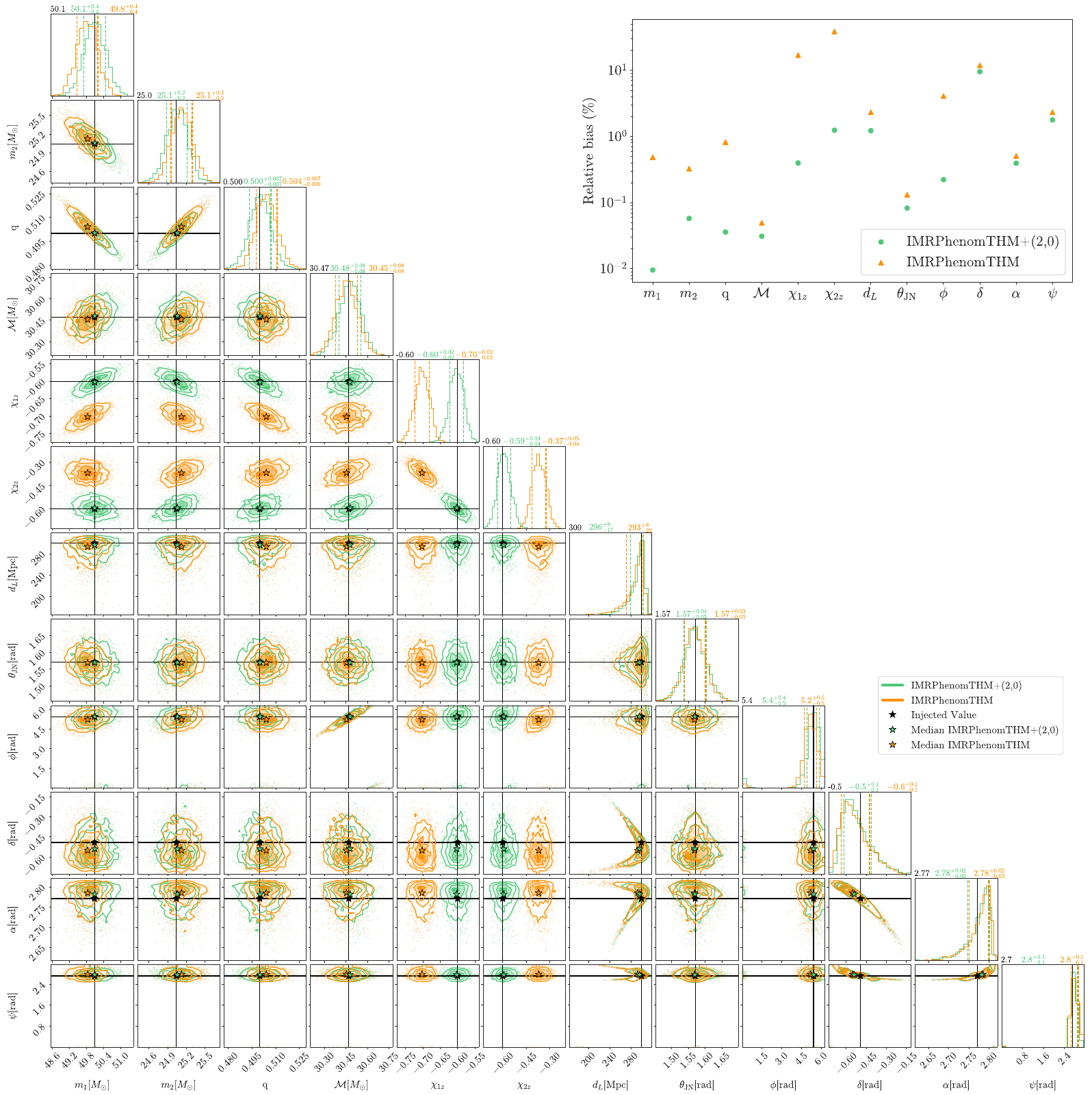}
\caption{Full corner plot and relative bias for Injection 8 in the LIGO network.}
\label{fig:inj8}
\end{figure}
\vspace*{\fill} 

\newpage
\section{Full corner plots for the injections in LIGO A$^{\#}$ and 3G networks}
\label{app:cornerplots_ET_CE}
In Figs.~\ref{fig:inj9CE}-\ref{fig:inj10ET}, we show the full corner plot of the posterior distributions of the sampled parameters for the injections performed with the networks of detectors combining LIGO A$^{\#}$ and ground-based 3G observatories. The format of the plots is the same as in Appendix~\ref{app:cornerplots}. In this case, the turquoise distributions correspond to the recovery with the model including the $(2,0)$ mode, while the pink show the recovery with the model neglecting this mode. Again, the top right panels in the figures present the relative bias in percentage for each of the sampled parameters and each version of the model. Turquoise dots stand for the model including the $(2,0)$ mode, while pink triangles represent the model neglecting this mode.

\begin{figure}[htp]
\includegraphics[width=1\textwidth]{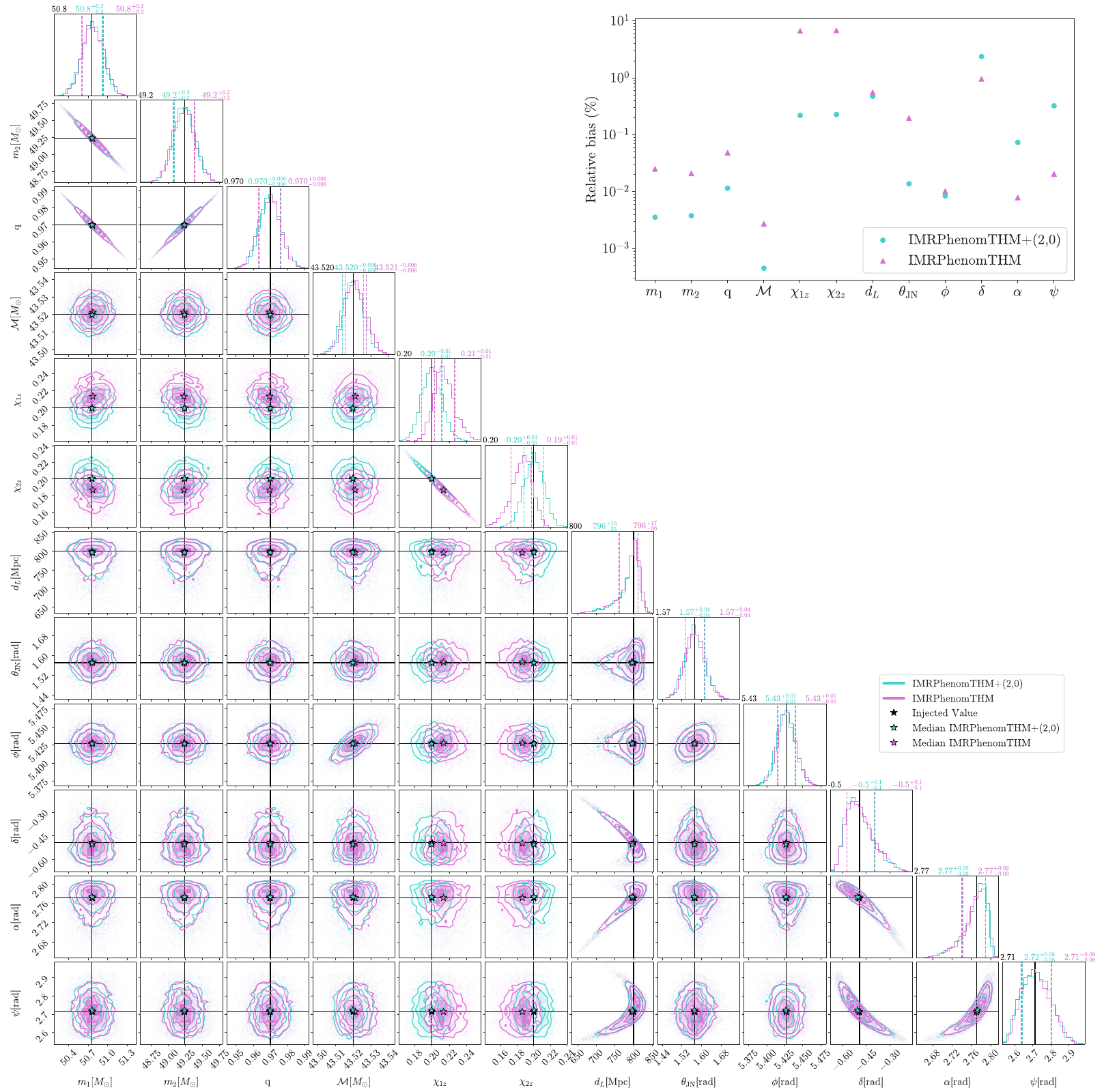}
    \caption{Full corner plot and relative bias for Injection 9 in the network composed of CE and L1.}
    \label{fig:inj9CE}
\end{figure}

\newpage
\thispagestyle{empty} 
\vspace*{\fill} 
\begin{figure}[htp]
\centering
\includegraphics[width=1\textwidth]{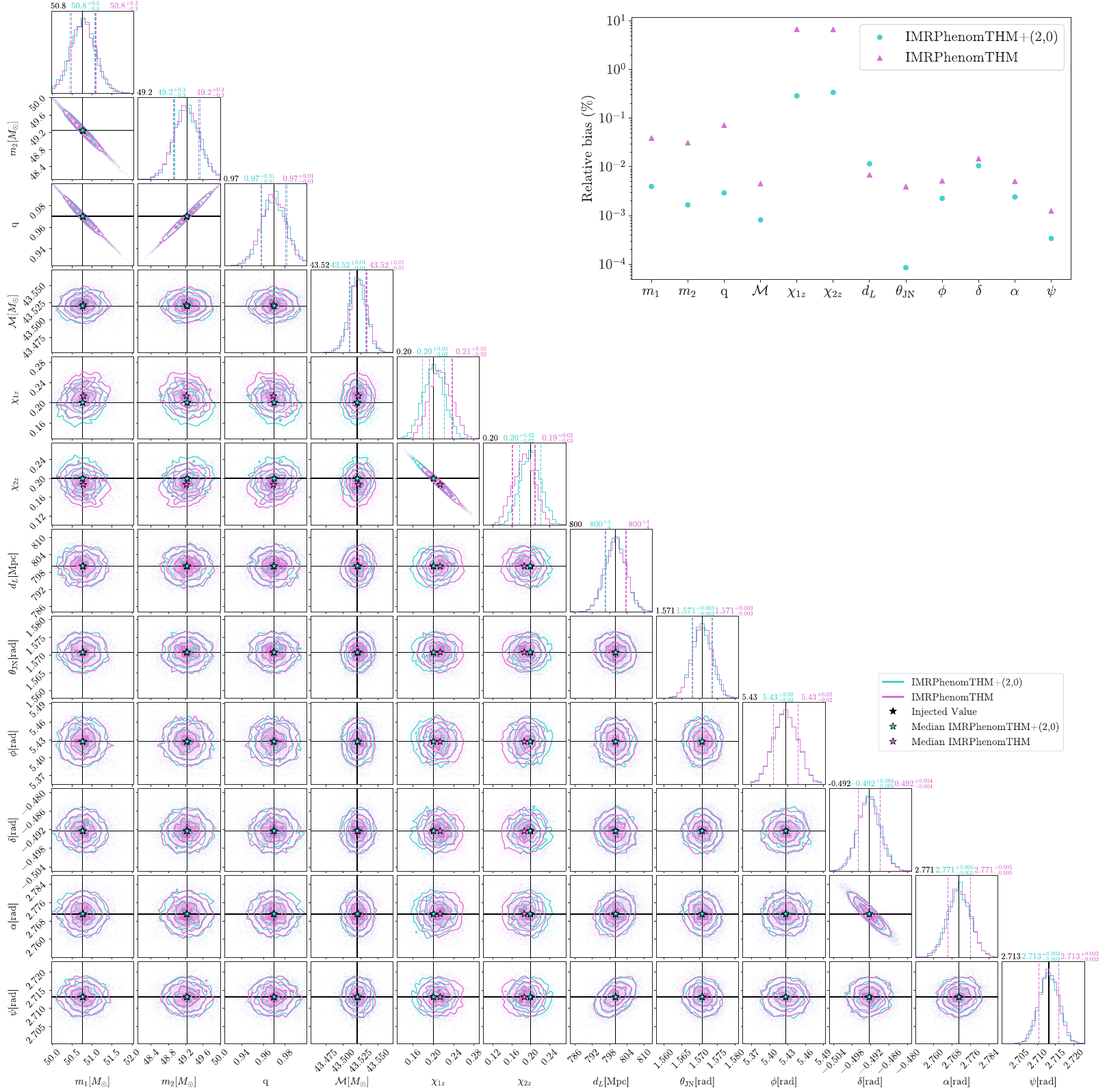}
\caption{Full corner plot and relative bias for Injection 9 in the network composed of ET, L1, and H1.}
\label{fig:inj9ET}
\end{figure}
\vspace*{\fill} 

\newpage
\thispagestyle{empty} 
\vspace*{\fill} 
\begin{figure}[htp]
\centering
\includegraphics[width=1\textwidth]{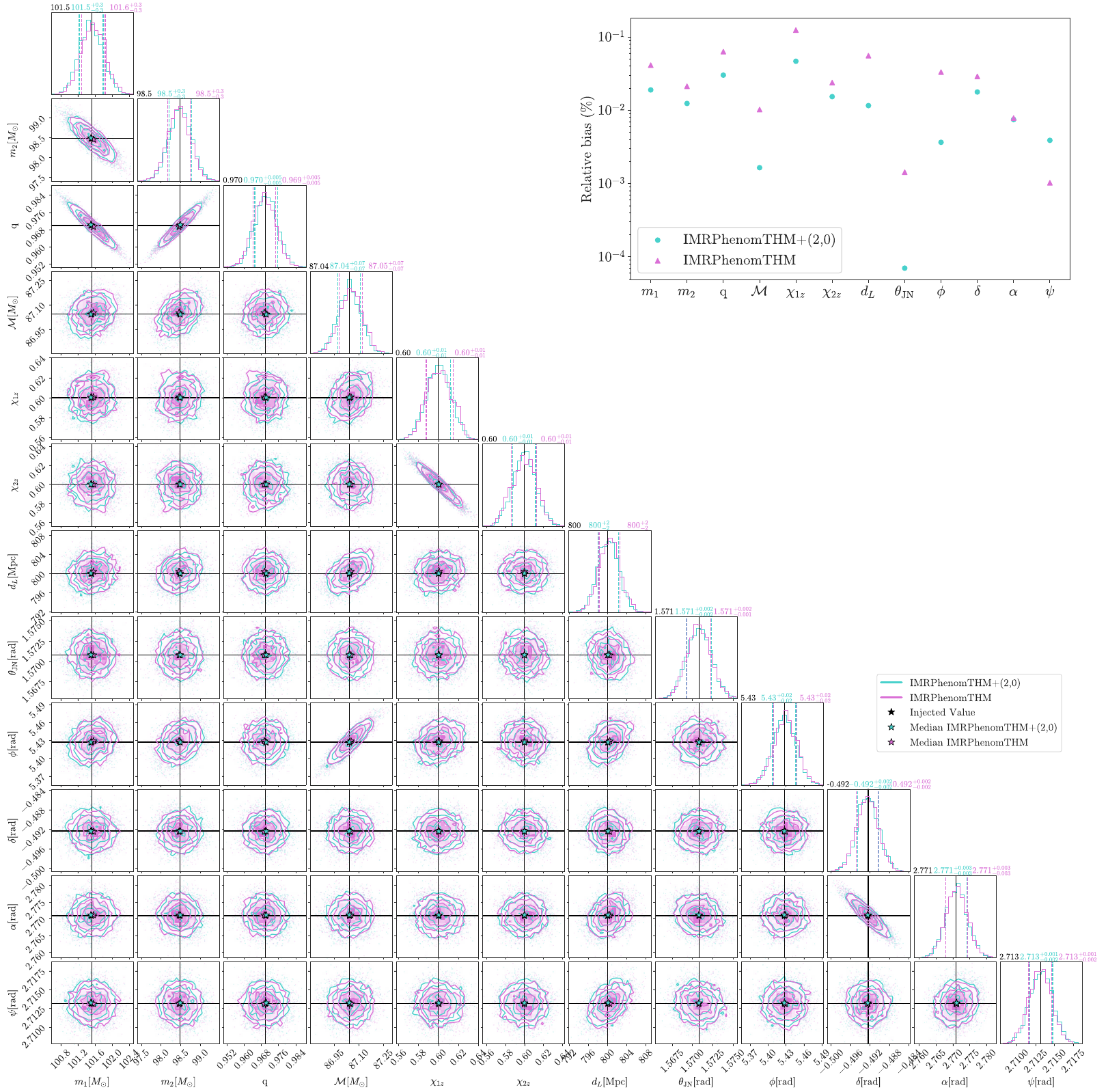}
\caption{Full corner plot and relative bias for Injection 10 in the network composed of ET, L1, and H1.}
\label{fig:inj10ET}
\end{figure}
\vspace*{\fill} 

\end{widetext}




\let\c\Originalcdefinition %
\let\d\Originalddefinition %
\let\i\Originalidefinition

\newpage
\bibliography{final_bibliography}

\end{document}